\documentstyle[sprocl]{article}

\input{psfig}
\def\Journal#1#2#3#4{{#1} {\bf #2}, #3 (#4)}

\def\NPB{{\em Nucl. Phys.} B}
\def\PLB{{\em Phys. Lett.}  B}
\def\PRL{\em Phys. Rev. Lett.}
\def\PRD{{\em Phys. Rev.} D}

\def\APJ{\em Astrophys. Journal}
\def\PL{\em Phys. Lett.}
\def\JETPL{\em JETP Lett.}
\def\SPSS{\em Sov. Phys. -Solid State}
\def\JPA{{\em J. Phys.} A}
\def\PR{\em Phys. Reports}
\def\APJL{\em Astrophys. J. Lett.}
\def\MNRAS{\em Mon. Not. Roy. Astr. Soc.}
\def\PR{\em Phys. Reports}
\def\CMP{\em Commun. Math. Phys.}
\def\N{\em Nature}
\def\PRSLA{{\em Proc. Roy. Soc. London} A}
\def\PTP{\em Prog. Theor. Phys.}
\def\IBID{\em ibid.}

\def\be{\begin{equation}}
\def\ee{\end{equation}}
\def\bea{\begin{eqnarray}}
\def\eea{\end{eqnarray}}

\begin{document}

\title{INFLATION}

\author{ G.LAZARIDES }

\address{Physics Division , School of Technology, 
Aristotle University of Thessaloniki,
\\ Thessaloniki GR 540 06, Greece.}


\maketitle\abstracts{
The shortcomings of the Standard Big Bang Cosmological Model 
as well as their resolution in the context of inflationary cosmology 
are discussed. The inflationary scenario and the subsequent 
oscillation and decay of the inflaton field are then studied 
in some detail. The density perturbations produced during inflation 
and their evolution during the matter dominated era are presented. 
The temperature fluctuations of the cosmic background radiation are 
summarized. The non-supersymmetric as well as the supersymmetric 
hybrid inflationary model is introduced 
and the `reheating' of the universe is analyzed in the context of 
the latter and a left-right symmetric gauge group. The scenario of 
baryogenesis via a primordial leptogenesis is considered in some detail. 
It is, finally, pointed out that, in the context
of a supersymmetric model based on a left-right symmetric gauge group, 
hybrid inflation, baryogenesis via primordial leptogenesis and neutrino 
oscillations are linked. This scheme, supplemented by a familiar ansatz 
for the neutrino Dirac masses and mixing of the two heaviest families 
and with the MSW resolution of the solar neutrino puzzle, implies that
$1\ {\rm {eV}\stackrel{_{<}}{_{\sim }}m_{\nu _{\tau }}
\stackrel{_{<}}{_{\sim }}9\ eV}$. The mixing angle 
$\theta_{\mu \tau }$ is predicted to lie in a narrow range which will 
be partially tested by the Chorus/Nomad experiment.}

\section{Shortcomings of the Big Bang Model}\label{sec:short}

The Standard Big Bang (SBB) Cosmological Model~\cite{wkt} has been 
very successful in explaining, among other things, the Hubble expansion 
of the universe, the existence of the Cosmic Background Radiation (CBR) 
and the abundances of the light elements which were formed during 
primordial nucleosynthesis. Despite its great successes, this model had 
a number of long-standing shortcomings which we will now summarize:

\subsection{Horizon Problem}\label{subsec:horizon}

The CBR, which we receive now, was emitted at the time of `decoupling' of 
matter and radiation (which essentially coincides with the time of 
recombination of atoms) when the cosmic temperature was 
$T_d  \approx 3,000~\rm{K}$. The decoupling time, $t_d$, can be 
calculated from
\begin{equation}
\frac {T_0}{T_d} = \frac {2.73~\rm{K}}{3,000~\rm{K}} = 
\frac {a (t_d)}{a(t_0)} = \left(\frac {t_d}{t_0}\right)^{2/3},
\label{eq:dec}
\end{equation}
where $t_0$, $T_0$ are  the present cosmic time and  temperature 
of CBR and $a(t)$ is the dimensionless scale factor of the universe at 
cosmic time $t$ normalized so that $a(t_0)=1$.
It turns out that $t_d \approx 200,000~h^{-1}$ years, where $h$ is 
the present value of the Hubble parameter in units of 
$100~\rm{km}~\rm{sec}^{-1}~\rm{Mpc}^{-1}$. 

\par
The distance over which the photons of the CBR have travelled 
since their emission is
\begin{equation}
a(t_0) \int^{t_{0}} _{t_{d}} \frac {dt^\prime}{a(t^\prime)} = 
3t_0\left[1 - \left(\frac {t_d}{t_0}\right)^{2/3}\right] 
\approx 3t_0 \approx 6,000~h^{-1}~\rm{Mpc}~,
\label{eq:lss}
\end{equation}
which essentially coincides with the present particle horizon size.
A sphere around us with radius equal to this distance is called the 
`last scattering surface' since the CBR observed now has been emitted 
from it. The particle horizon size at $t_d$ was $2H^{-1} (t_d) = 
3t_d \approx 0.168~h^{-1}~\rm{Mpc}$~($H(t)$ being the Hubble 
parameter at cosmic time $t$) and expanded till now to become 
$0.168~h^{-1} (a(t_0)/a(t_d))~{\rm{Mpc}}\approx 184~h^{-1}$ Mpc.
The angle subtended by this `decoupling' horizon at present is
$\theta_{d} \approx 184/6,000 \approx 0.03~\rm{rads} \approx 2~^o$. 
Thus, the sky splits into $4 \pi/(0.03)^2 \approx 14,000$ patches 
that never communicated causally before sending light to us. The question 
then arises how come the temperature of the black body radiation from 
all these patches is so accurately tuned as the measurements of the 
Cosmic Background Explorer~\cite{cobe} (COBE) require
($\delta T/T \approx 6.6 \times 10^{-6}$).

\subsection{Flatness Problem} \label{subsec:flatness}

The present energy density, $\rho$, of the universe has been observed 
to lie in the relatively narrow range $0.1 \rho_c \stackrel{_{<}}
{_{\sim }}\rho \stackrel{_{<}}{_{\sim }}2 \rho_c$, where 
$\rho_c$ is the critical energy density corresponding to a flat 
universe. The lower bound has been derived from estimates of galactic 
masses using the virial theorem whereas the upper bound from the volume 
expansion rate implied by the behavior of
galactic number density at large distances. The Friedmann equation
\begin{equation}
H^2 = \frac {8 \pi G} {3} \rho - \frac {k} {a^2} = 
\frac {8 \pi G}{3} \rho_{c}~,
\label{eq:fried}
\end{equation}
where $H=\dot{a}(t)/a(t)$ (overdots denote derivation with respect to 
cosmic time) is the Hubble parameter, $G$ the Newton' s constant and $k$ 
negative zero or positive for an open, flat or closed
universe respectively, implies that $(\rho - \rho_c)/\rho_c =
3 (8 \pi G \rho_c)^{-1}(k/a^2)$ is proportional to $a$, for matter 
dominated universe. Consequently, in the early universe, 
$ |(\rho - \rho_c)/\rho_c|\ll 1$ and the question  arises why 
the initial energy density of the universe was so finely tuned to be 
equal to its critical value.

\subsection{Magnetic Monopole Problem} \label {subsec:monopole}

This problem arises only if we combine the SBB Model with Grand Unified 
Theories~\cite{ggps} (GUTs) of strong, weak and electromagnetic 
interactions. In accordance with GUTs, the universe underwent~\cite{kl} 
a phase transition during which the GUT gauge symmetry group, $G$, 
broke down to the Standard Model gauge group, $G_S$. This breaking was 
due to the fact that, at a critical temperature $T_c$, an appropriate 
higgs field, $\phi$, developed a non-zero vacuum
expectation value (vev). Assuming that this phase transition
was a second order one, we have $\langle \phi\rangle(T) 
\approx \langle \phi\rangle(T=0)( 1 - T^2/T^2_c)^{1/2}$, $m_H (T)
\approx \lambda \langle \phi\rangle (T)$, for the temperature 
dependent vev and mass of the higgs field respectively at $T \leq T_c$ 
($\lambda$ is an appropriate higgs coupling constant).

\par
The GUT phase transition produces magnetic monopoles~\cite{monopole} 
which are localized deviations from the vacuum with radius 
$\sim M_X^{-1}$, 
energy $\sim M_X/\alpha_G$ and $\phi =0$ at their center ($M_X$  is 
the GUT mass scale and $\alpha_G = g^2_{G}/4\pi$ with $g_G$ being the 
GUT gauge coupling constant). The vev of the higgs field on a sphere, 
$S^2$, with radius $\gg M_{X}^{-1}$
around the monopole lies on the vacuum manifold $G/G_S$ and we, thus, 
obtain a  mapping: $S^2\longrightarrow G/G_S$. If this mapping is 
homotopically non-trivial the topological stability of the magnetic 
monopole is guaranteed.

\par
Monopoles can be produced when the fluctuations of $\phi$ over $\phi=0$ 
between the vacua at $\pm \langle \phi\rangle(T)$ cease to be frequent. 
This happens when the free energy needed for $\phi$ to fluctuate from 
$\langle \phi\rangle (T)$ to zero in a region of radius equal to the 
higgs correlation length $\xi(T) = m^{-1}_H (T)$  exceeds $T$. This 
condition reads $(4\pi/3) \xi^3 \Delta V \stackrel{_{>}}
{_{\sim }} T$, where $\Delta V \sim \lambda^2 
\langle \phi\rangle^4$ is the difference in free energy 
density between $\phi =0$ and $\phi=\langle \phi\rangle(T)$. The 
Ginzburg temperature~\cite{ginzburg}, $T_G$, corresponds to the
saturation of this inequality. So, at 
$T \stackrel{_{<}}{_{\sim }} T_G$, the fluctuations over
$\phi=0$ stop and $\langle \phi\rangle$ settles on the vacuum manifold 
$G/G_S$. At $T_G$, the universe splits into regions of size $\xi_G \sim 
(\lambda^2 T_c)^{-1}$, the higgs correlation length at $T_G$, with the 
higgs field more or less aligned in each region. Monopoles are produced
at the corners where such regions meet (Kibble~\cite{kibble} mechanism) 
and their number density is estimated to be 
$n_M \sim {\rm{p}} \xi_{G}^{-3} \sim {\rm{p}} \lambda^{6} T_{c}^{3}$, 
where $\rm{p} \sim \rm{1/10}$ is a geometric factor. The `relative' 
monopole number density then turns out to  be
$r_M =n_M/T^3 \sim \rm{10^{-6}}$. We can derive a lower bound on 
$r_M$ by pure causality considerations. The higgs field $\phi$ 
cannot be correlated at distances bigger than the particle horizon size, 
$2t_G$, at $T_G$. This gives the causality bound
\begin{equation}
n_M  \stackrel{_{>}}{_{\sim }}\frac {\rm{p}} 
{\frac{4 \pi}{3}(2t_G)^3}~,
\label{eq:causal}
\end{equation}
which implies that $r_M\stackrel{_{>}}{_{\sim }}\rm{10^{-10}}$.

\par
The subsequent evolution of monopoles, after $T_G$, is governed by the
equation~\cite{preskill} (overdots denote derivation with respect to 
cosmic time)
\begin{equation}
\frac {dn_M}{dt} = - D n_{M}^{2} - 3 \frac {\dot{a}}{a} n_{M}~,
\label{eq:evol}
\end{equation}
where the first term in the right hand side (with $D$ being an 
appropriate constant) describes the dilution of monopoles due to 
monopole-antimonopole annihilation while the second term corresponds to 
their dilution by the cosmological expansion. The monopole-antimonopole
annihilation proceeds as follows. Monopoles diffuse towards antimonopoles
through the plasma of charged particles, capture each other in Bohr 
orbits and eventually annihilate. The annihilation is effective 
provided the mean free path of the monopoles in the plasma of charged 
particles does not exceed their capture distance. This happens at cosmic 
temperatures $T \stackrel{_{>}}{_{\sim }} \rm{10^{12}~GeV}$. 
The overall result is that, if the initial relative magnetic monopole
density $r_{M,\rm{in}} \stackrel{_{>}}{_{\sim }}\rm{10^{-9}} 
( \stackrel{_{<}}{_{\sim }}\rm{10^{-9}})$, the final one
$r_{M,\rm{fin}} \sim 10^{-9} ( \sim r_{M,\rm{in}})$. This combined 
with the causality bound yields $r_{M,\rm{fin}} 
\stackrel{_{>}}{_{\sim }}\rm{10^{-10}}$. 
However, the requirement that monopoles do not dominate the
energy density of the universe at nucleosynthesis gives
\begin{equation}
r_M (T \approx 1~\rm{MeV}) \stackrel{_{<}}
{_{\sim }}\rm{10^{-19}}~,
\label{eq:nucleo}
\end{equation}
and we have a discrepancy of about ten orders of magnitude.

\subsection{Density Fluctuations}\label{subsec:fluct}

For structure formation~\cite{structure} in the universe, we need a 
primordial density perturbation, $\delta \rho/ \rho$, at all scales 
with a nearly flat spectrum~\cite{hz}. We also need some explanation 
of the temperature fluctuations, $\delta T/T$, of CBR observed by 
COBE~\cite{cobe} at angles $\theta \stackrel{_{>}}{_{\sim }} 
\theta_d \approx 2~^o$ which violate causality 
(see Sec.\ref{subsec:horizon}). 

\par
Let us expand $\delta \rho/\rho$ in plane waves
\begin{equation}
\frac {\delta \rho} {\rho} (\bar{x},t) = 
\int d^3 k\delta_{\bar{k}}(t)e^{i\bar{k} \bar{x}}~, 
\label{eq:plane}
\end{equation}
where $\bar{x}$ is a comoving vector in 3-space and $\bar{k}$ is 
the comoving wave vector with $k=|\bar{k}|$ 
being the comoving wave number ($\lambda=2 \pi/k$ is the
comoving wave length whereas the physical wave length is 
$ \lambda _{\rm{phys}} =a(t) \lambda$).
For $\lambda_{\rm{phys}} \leq H^{-1}$, the time evolution of 
$\delta_{\bar{k}}$ is described by the Newtonian equation
\begin{equation}
\ddot{\delta}_{\bar{k}} + 2 H \dot {\delta}_{\bar{k}} + 
\frac {v_{s}^{2}k^2}{a^2}\delta_{\bar{k}}=
4 \pi G \rho \delta_{\bar{k}}~,
\label{eq:newton}
\end{equation}
where the second term in the left hand side comes from the cosmological
expansion and the third is the `pressure' term ($v_s$ is the velocity of 
sound given by $v^{2}_{s}=dp/d\rho$, where $p$ is the mean pressure). 
The right hand side of this equation corresponds to the gravitational 
attraction.

\par
For the moment, let us put $H$=0 (static universe). In this case, 
there exists a characteristic wave number 
$k_J$, the Jeans wave number, given by $k^{2}_{J}=4 \pi G a^{2} \rho/
v^{2}_{s}$ and having the following property.
For $k \geq k_J$,  pressure dominates over gravitational attraction 
and the density perturbations just oscillate, whereas, for $k \leq k_J$, 
gravitational attraction dominates and the density perturbations grow 
exponentially. In particular, for $p$=0 (cold dark matter), 
$v_s=0$ and all scales are Jeans unstable with
\begin{equation}
\delta_{\bar{k}} \propto {\rm{exp}}(t/\tau)~,~\tau=
(4 \pi G \rho)^{-1/2}~.
\label{eq:jeans}
\end{equation}

\par
Now let us take $H\neq 0$. Since the cosmological expansion pulls the
particles apart, we get a smaller growth:
\begin{equation}
\delta_{\bar{k}}\propto a(t) \propto t^{2/3}~,
\label{eq:growth}
\end{equation}
in the matter dominated case. For a radiation dominated universe 
($p \neq 0$), we get essentially no growth of the density 
perturbations. This means that, in order to have structure formation 
in the universe, which requires $\delta \rho/\rho \sim 1$, we must have
\begin{equation}
(\frac {\delta \rho} {\rho})_{\rm{equ}}\sim 
4 \times 10^{-5} ( \Omega_0 h)^{-2}~,
\label{eq:equi}
\end{equation}
at the equidensity point (where the energy densities of matter and 
radiation coincide), since the available growth factor for 
perturbations is given by $a_0/a_{\rm{equ}} 
\sim 2.5 \times 10^4 (\Omega_0 h)^2$. Here $\Omega_0=
\rho_0/\rho_c$, where $\rho_0$ is the present energy density of 
the universe. The question then is where these primordial density 
fluctuations originate from.

\section{Inflation}\label{sec:inflation}

Inflation~\cite{guth,lindebook} is an idea which solves simultaneously 
all four cosmological puzzles and can be summarized as follows. 
Suppose there is a real scalar 
field $\phi$ (the inflaton) with (symmetric) potential energy density 
$V(\phi)$ which is quite `flat' near $\phi=0$ and has minima at 
$\phi = \pm\langle \phi\rangle$ with 
$V(\pm \langle \phi\rangle)=0$. 
At high enough $T$' s, $\phi =0$ in the universe due to the temperature 
corrections in $V(\phi)$. As $T$ drops, the effective potential density 
approaches the $T$=0 potential but a little potential barrier separating 
the local minimum at $\phi=0$ and the vacua at 
$\phi = \pm\langle \phi\rangle$ still remains. At some point, 
$\phi$ tunnels out to $\phi_1 \ll\langle \phi\rangle$ and a 
bubble with $\phi=\phi_1$ is created in the universe. The field then 
rolls over to the minimum of $V(\phi)$ very slowly (due to the flatness 
of the potential). During this slow roll-over, the energy density 
$\rho \approx V(\phi=0) \equiv V_0$ remains essentially constant for 
quite some time. The Lagrangian density
\begin{equation}
L=\frac{1}{2} \partial_{\mu} \phi \partial^{\mu} \phi - V(\phi)
\label{eq:lagrange}
\end{equation}
gives the energy-momentum tensor
\begin{equation}
T_{\mu}~^{\nu} = - \partial_\mu \phi \partial^\nu \phi + 
\delta_{\mu}~^{\nu}\left(\frac{1}{2}
\partial_\lambda \phi \partial^\lambda \phi - V (\phi)\right)~,
\label{eq:energymom}
\end{equation}
which during the slow roll-over takes the form $T_{\mu}~^{\nu} 
\approx - V_{0}~\delta_{\mu}~^{\nu}$.
This means that $\rho \approx -p \approx V_0$, i.e., the pressure 
$p$ is negative and equal in magnitude with the energy density $\rho$, 
which is consistent with the continuity equation $\dot{\rho} = 
-3H(\rho + p)$. Since, as we will see, $a(t)$ grows very fast, 
the `curvature' term, $k/a^2$, in Eq.\ref{eq:fried} becomes 
subdominant and we get
\begin{equation}
H^2 \equiv \left(\frac {\dot{a}}{a}\right)^2 = \frac {8 \pi G} {3} V_0~,
\label{eq:inf}
\end{equation}
which gives $a(t) \propto e^{Ht},~H^2 =(8 \pi G/3)V_0$ =~constant.
So the bubble expands exponentially for some time and $a(t)$ grows by 
a factor
\begin{equation}
\frac {a(t_f)}{a(t_i)} ={\rm{exp}} H(t_f - t_i) 
\equiv {\rm{exp}} H \tau~,
\label{eq:efold}
\end{equation}
between an initial ($t_i$) and a final ($t_f$) cosmic time.

\par
The inflationary scheme just described, which is known as the 
new~\cite{new}
inflationary scenario (with
the inflaton field starting from the origin, $\phi$=0), is certainly
not the only realization of the idea of inflation. Another interesting 
possibility is to consider the universe as it
emerges at the Planck time $t_{P}=m_{P}^{-1}, m_{P} =
M_{P} / \sqrt{8 \pi}$ with $ M_{P} = 1.22 \times 10^{19} $ GeV, 
where the fluctuations of gravity cease to exist. We can imagine a 
region of size $\ell_{P} \approx m_{P}^{-1}$ where the inflaton 
field acquires a large and almost uniform value and carries 
negligible kinetic energy. 
Under certain circumstances this region can inflate (exponentially 
expand) as $\phi$ rolls down towards its vacuum value. This type of 
inflation with the inflaton starting from large values is known as the 
chaotic~\cite{chaotic} inflationary scenario.

\par
We will now show that, with an adequate number of e-foldings, 
$N=H \tau$, the first three cosmological puzzles are easily 
resolved (we will leave the question of density perturbations for later).

\subsection{Resolution of the Horizon Problem} \label{subsec:infhor}

The particle horizon during inflation (exponential expansion)
\begin{equation}
d(t) = e^{Ht} \int^t_{t_{i}} \frac {d t^\prime}{e^{Ht^\prime}} 
\approx H^{-1}{\rm{exp}}H(t-t_i)~,
\label{eq:horizon}
\end{equation}
for $t-t_i \gg H^{-1}$, grows as fast as $a(t)$. At the end of 
inflation ($t=t_f$),~$d(t_f) \approx H^{-1}{\rm{exp}} H \tau$ and 
the field $\phi$ starts oscillating about the minimun of the 
potential at $\phi = \langle \phi\rangle$. It then decays and 
`reheats'~\cite{reheat} the universe at a temperature 
$T_r \sim 10^9~{\rm{GeV}}$ (see Sec.\ref{sec:reheathybrid}). 
The universe, after that, goes back to normal big bang cosmology. 
The horizon $d(t_{f})$ is stretched during the period of 
$\phi$-oscillations by some factor $\sim 10^9$ depending
on  details and between $T_r$ and the present era by a factor 
$T_r/T_0$. So it finally becomes equal to 
$H^{-1} e^{H \tau} 10^9 (T_r/T_0)$, which should exceed 
$2H_{0}^{-1}$ in order to solve the horizon 
problem. Taking $V_0 \approx M_{X}^{4},~M_{X} \sim 10^{16}$ GeV, 
we see that, with $N = H \tau\stackrel{_{>}}{_{\sim }} 55$, 
the horizon problem is evaded.

\subsection{Resolution of the Flatness Problem}\label{subsec:infflat}

The curvature term of the Friedmann equation, at present, is given by
\begin{equation}
\frac {k}{a^2} \approx \left(\frac {k}{a^2}\right)_{bi}  
e^{-2H \tau}~10^{-18} \left(\frac {10^{-13}~{\rm{GeV}}}
{10^9~{\rm{ GeV}}}\right)^2, 
\label{eq:curvature}
\end{equation}
where the terms in the right hand side correspond to the curvature 
term before inflation, and its growth factors during inflation, during 
$\phi$ -oscillations and after `reheating' respectively. Assuming 
$(k/a^2)_{bi} \sim (8 \pi G/3) \rho \sim H ^2$~~
$(\rho \approx V_0)$, we get $k/a_{0}^{2} H_{0}^{2} 
\sim 10^{48}~e^{-2H \tau}$ which gives $(\rho_0 - \rho_c)/\rho_c 
\equiv \Omega_0 - 1 = k/a_{0}^{2} H_{0}^{2} \ll 1$, for 
$H \tau \gg 55$. In fact, strong inflation implies that the present 
universe is flat with a great accuracy.

\subsection{Resolution of the Monopole Problem}\label{subsec:infmono}

It is obvious that, with a number of e-foldings $\stackrel{_{>}}
{_{\sim }} 55$, the primordial monopole density is diluted by at 
least 70 orders of magnitude and they become totally irrelevant. 
Also, since $T_r \ll m_M$, there is no production of magnetic monopoles 
after `reheating'.

\section{Detailed Analysis of Inflation} \label{sec:detail}

The Hubble parameter is not exactly constant during inflation as we,
naively, assumed so far. It actually depends on the value of $\phi$:
\begin{equation}
H^{2}(\phi) = \frac {8 \pi G} {3} V (\phi)~.
\label{eq:hubble}
\end{equation}
To find the evolution equation for $\phi$ during inflation, we vary the 
action
\begin{equation}
S= \int \sqrt{-{\rm{det}}(g)}~ d^{4}x \left(\frac {1}{2} 
\partial_ {\mu} \phi \partial^{\mu} \phi - V(\phi) + 
M(\phi)\right)~,
\label{eq:action}
\end{equation}
where $g$ is the metric tensor and $M(\phi)$ represents the coupling 
of $\phi$ to `light' matter causing its decay. We find
\begin{equation}
\ddot{\phi} + 3H \dot{\phi} + \Gamma_{\phi} \dot{\phi} + 
V^{\prime}(\phi) = 0~,
\label{eq:evolution}
\end{equation}
where the prime denotes derivation with respect to $\phi$ and 
$\Gamma_{\phi}$ is the decay width~\cite{width} of the inflaton. 
Assume, for the moment, that the decay time
of $\phi $, $t_d = \Gamma_{\phi}^{-1}$, is much greater than 
$H^{-1}$, the expansion time for inflation.
Then the term $\Gamma_{\phi} \dot{\phi}$ can be ignored and 
Eq.\ref{eq:evolution} reduces to
\begin{equation}
\ddot{\phi} + 3 H \dot{\phi} + V^{\prime}(\phi) = 0~.
\label{eq:reduce}
\end{equation}
Inflation is by definition the situation where $\ddot{\phi}$ is 
subdominant to the `friction' term $3H \dot{\phi}$ in this equation 
(and the kinetic energy density is subdominant to the potential energy 
density). Eq.\ref{eq:reduce} then
further reduces to the inflationary equation~\cite{slowroll}
\begin{equation}
3H \dot{\phi} = - V^{\prime} (\phi)~,
\label{eq:infeq}
\end{equation}
which gives
\begin{equation}
\ddot{\phi} = - \frac {V^{\prime\prime}(\phi)\dot{\phi}}
{3H(\phi)} + \frac {V^{\prime}(\phi)}
{3H^{2}(\phi)} H^\prime (\phi) \dot{\phi}~.
\label{eq:phidd}
\end{equation}
Comparing the two terms in the right hand side of this equation with 
the `friction' term in Eq.\ref{eq:reduce}, we obtain the conditions 
for inflation (slow roll conditions):
\begin{equation}
\eta \equiv \frac{M_{P}^{2}}{8 \pi} \bigg | 
\frac {V^{\prime\prime}(\phi)}
{V(\phi)} \bigg | \leq 1~,
~\epsilon \equiv \frac {M_{P}^{2}}{16 \pi} 
\left(\frac {V^{\prime}(\phi)}
{V(\phi)}\right)^{2} \leq 1~.
\label{eq:src}
\end{equation}
The end of the slow `roll-over' occurs when either of the
these inequalities is saturated. If
$\phi_f$ is the value of $\phi$ at the end of inflation, then
$t_f \sim H^{-1}(\phi_f)$.

\par
The number of e-foldings during inflation can be calculated as
follows:
\begin{equation}
N(\phi_{i}\rightarrow \phi_{f}) \equiv \ell n 
\left(\frac {a(t_{f})}
{a(t_{i})}\right) = \int^{t_{f}} _{t_{i}} Hdt
= \int^{\phi_{f}}_{\phi_{i}} 
\frac {H (\phi)}{\dot{\phi}} d \phi = -
\int^{\phi_{f}}_{\phi_{i}} \frac {3 H^2 (\phi) d \phi}
{V^{\prime}(\phi)}~,
\label{eq:nefolds}
\end{equation}
where  Eqs.\ref{eq:efold}, \ref{eq:infeq} and the definition of
$H = \dot{a}/a$ were used. For simplicity, we can
shift the field $\phi$ so that the global minimum of the potential is
displaced at $\phi$ = 0. Then, if $V(\phi) = \lambda \phi^{\nu}$
during inflation, we have
\begin{equation}
N(\phi_{i} \rightarrow \phi_{f}) = 
- \int^{\phi_{f}}_{\phi_{i}} \frac
{3H^2(\phi)d\phi}{V^{\prime}(\phi)} = 
- 8 \pi G \int^{\phi_{f}}_{\phi_{i}} \frac
{V(\phi)d\phi}{V^{\prime}(\phi)}=\frac {4 \pi G}{\nu}
(\phi^{2}_{i}-\phi^{2}_{f})~.
\label{eq:expefold}
\end{equation}
Assuming that $\phi_{i} \gg \phi_{f}$, this reduces to $N(\phi) = 
(4 \pi G/\nu)\phi^2$.

\section{Coherent Field Oscillations}\label{sec:osci}

After the end of inflation at cosmic time $t_f$, the term 
$\ddot{\phi}$ takes over and Eq.\ref{eq:reduce} reduces to
$\ddot{\phi} + V^{\prime}(\phi)=0$, which means that $\phi$ 
starts oscillating coherently about the global minimum of the potential. 
In reality, due to the `friction' term, $\phi$ performs damped 
oscillations with a rate of energy density loss given by
\begin{equation}
\frac{d \rho}{dt} = \frac {d}{dt}\left(\frac{1}{2} 
\dot{\phi}^2 + V(\phi)\right)
= - 3H \dot{\phi}^2=-3H(\rho+p)~,
\label{eq:damp}
\end{equation}
where $\rho = \dot{\phi}^2/2 + V(\phi) $ and the pressure $p = 
\dot{\phi}^2/2 - V(\phi)$.
Averaging $p$ over one oscillation of
$\phi$, we write~\cite{oscillation} $\rho+p=\gamma \rho$. 
Eq.\ref{eq:damp} then becomes $\dot{\rho} = - 3 H \gamma \rho$, 
which gives $d \rho/\rho = -3 \gamma da/a$ and
$\rho \propto a^{-3 \gamma}$. The Friedmann equation then takes the 
form $\dot{a}/a \propto a^{-3 \gamma/2}$ and we obtain $a(t) 
\propto t^{2/3 \gamma}$. 

\par
The number $\gamma$ for an oscillating 
field can be written as (assuming a symmetric potential)
\begin{equation}
\gamma = \frac {\int^{T}_{0} \dot{\phi}^{2} dt}
{\int^{T}_{0} \rho dt} =
\frac{\int^{\phi_{{\rm{max}}}}_{0} \dot{\phi}d \phi} 
{\int^{\phi_{{\rm{max}}}}_{0}(\rho/\dot{\phi}) d\phi}~,
\label{eq:gamma}
\end{equation}
where $T$ and $\phi_{{\rm{max}}}$ are the period and the amplitude 
of the oscillation respectively. From the equation 
$\rho = \dot{\phi}^2/2 + V(\phi)=V_{{\rm{max}}}$, 
where $V_{{\rm{max}}}$ is the maximal potential energy density, 
we obtain $\dot{\phi} = \sqrt{2(V_{{\rm{max}}} - V(\phi))}$. 
Substituting this in Eq.\ref{eq:gamma} we get~\cite{oscillation}
\begin{equation}
\gamma = \frac {2 \int^{\phi_{{\rm{max}}}}_{0} 
(1-V/V_{{\rm{max}}})^{1/2}
d\phi}{\int^{\phi_{{\rm{max}}}}_{0} 
(1-V/V_{{\rm{max}}})^{-1/2} d \phi}~~\cdot
\label{eq:gammafinal}
\end{equation}
For a potential of the simple form $V(\phi) = 
\lambda \phi^{\nu}$,~$\gamma$ is
readily found to be given by $\gamma=2 \nu/(\nu +2)$. Consequently, 
in this case, $\rho \propto a^{-6\nu/(\nu+2)}$ and 
$a(t) \propto t^{(\nu+2)/3 \nu}$. For $\nu=2$, in particular, 
one has $\gamma$=1,~$\rho \propto a^{-3}$,~$a(t)
\propto t^{2/3}$ and the oscillating field behaves like 
pressureless `matter'. This is not unexpected since a 
coherent oscillating massive free field corresponds to a distribution 
of static massive particles. For $\nu$=4, however, we obtain 
$\gamma = 4/3$,~$\rho 
\propto a^{-4}$,~$a(t) \propto t^{1/2}$ and the system resembles 
`radiation'. For $\nu = 6$, one has $\gamma=3/2$,~$ \rho 
\propto a^{-4.5}$,~$a(t)\propto t^{4/9}$
and the expansion is slower than in a `radiation' dominated universe 
(the pressure is higher than in `radiation').

\section{Decay of the Field $\phi$}\label{sec:decay}

Reintroducing the `decay' term $\Gamma_{\phi} \dot{\phi}$, 
Eq.\ref{eq:evolution} can be written as
\begin{equation}
\dot{\rho} = \frac{d}{dt} \left(\frac{1}{2} \dot{\phi}^2 + 
V(\phi)\right) = - (3H + \Gamma_\phi)\dot{\phi}^2~, 
\label{eq:decay}
\end{equation}
which is solved~\cite{reheat,oscillation} by
\begin{equation}
\rho(t)=\rho_{f}\left(\frac{a(t)}{a(t_{f})}\right)^{-3 \gamma} 
{\rm{exp}} [ -\gamma \Gamma_{\phi}(t-t_f)]~, 
\label{eq:rho}
\end{equation}
where $\rho_f$ is the energy density at the end of inflation at cosmic 
time $t_f$. The second and third factors in the right hand side of this 
equation represent the dilution of the field energy due to the expansion 
of the universe and the decay of $\phi$ to light particles respectively.

\par 
All pre-existing `radiation' (known as `old radiation') was diluted by 
inflation, so the only `radiation' present is the one produced by the 
decay of $\phi$ and is known as `new radiation'. Its energy density satisfies
~\cite{reheat,oscillation} the equation
\begin{equation}
\dot{\rho}_{r} = - 4 H \rho_{r} + \gamma \Gamma_{\phi} \rho~,
\label{eq:newrad}
\end{equation}
where the first term in the right hand side represents the dilution of 
radiation due to the cosmological expansion while the second one is 
the energy density transfer from $\phi$ to `radiation'. Taking 
$\rho_{r}(t_f)$=0, this equation gives~\cite{reheat,oscillation}
\begin{equation}
\rho_{r}(t) = \rho_{f}\left(\frac {a(t)}
{a(t_{f})}\right)^{-4} \int^{t}_{t_{f}} 
\left(\frac{a(t^{\prime})}{a(t_{f})}\right)^{4-3 \gamma}
e^{ -\gamma \Gamma_{\phi} (t^{\prime}-t_f)} 
~\gamma \Gamma_{\phi} dt^{\prime}~.
\label{eq:rad}
\end{equation}
For $t_{f} \ll t_{d}$ and $\nu =2$, this expression is approximated 
by
\begin{equation}
\rho_{r}(t)=\rho_{f}\left(\frac {t}{t_f}\right)^{-8/3} 
\int^{t}_{0} 
\left(\frac{t^{\prime}}{t_{f}}\right)^{2/3}
e^{-\Gamma_{\phi}t^{\prime}} dt^{\prime}~,
\label{eq:appr}
\end{equation}
which, using the formula
\begin{equation}
\int_{0}^{u} x^{p-1} e^{-x}dx = e^{-u}~\sum^{\infty}_{k=0}~ 
\frac {u^{p+k}}{p(p+1)\cdot\cdot\cdot(p+k)}~~,
\label{eq:formula}
\end{equation}
can be written as
\begin{equation}
\rho_{r} = \frac {3}{5}~\rho~\Gamma_{\phi}t 
\left[1 + \frac {3}{8}~\Gamma_{\phi}t
+ \frac {9}{88}~(\Gamma_{\phi}t)^2+ \cdots \right]~,
\label{eq:expand}
\end{equation}
with $\rho = \rho_{f} (t/t_{f})^{-2}{\rm{exp}}
(-\Gamma_{\phi}t)$ being the energy density of the field $\phi$ 
which performs damped oscillations and decays into `light' particles. 

\par
The energy density of the `new radiation' 
grows relative to the energy density of the oscillating field and becomes 
essentially equal to it at a cosmic time $t_{d} = \Gamma_{\phi}^{-1}$ as 
one can deduce from Eq.\ref{eq:expand}. After this time, the universe 
enters  into the radiation dominated era and the normal big bang cosmology 
is recovered. The temperature at $t_{d},~T_{r}(t_{d})$, 
is historically called the `reheating' temperature although no supercooling 
and subsequent reheating of the universe actually takes place. Using the 
time to temperature relation for a radiation dominated universe we find 
that
\begin{equation}
T_{r} = \left(\frac {45}{16 \pi^{3}g_*}\right)^{1/4} 
(\Gamma_{\phi} M_{P})^{1/2}~,
\label{eq:reheat}
\end{equation}
where $g_*$ is the effective number of degrees of freedom. For a
potential of the type $V(\phi)=\lambda\phi^{\nu}$, the total 
expansion of the universe during the period of damped field 
oscillations is
\begin{equation}
\frac{a(t_{d})}{a(t_{f})} = \left(\frac{t_{d}}{t_{f}}\right)^
{\frac{\nu + 2}{3\nu}}~.
\label{eq:expansion}
\end{equation}

\section{Density Perturbations}\label{sec:density}

We are now ready to sketch how inflation solves the density 
fluctuation problem described in Sec.\ref{subsec:fluct}. As a matter 
of fact, inflation not only homogenizes the universe but also provides 
us with the primordial density fluctuations necessary for the structure 
formation in the universe. To understand the origin of these fluctuations, 
we must first introduce the notion of `event horizon'. Our `event horizon', 
at a cosmic time $t$, includes all points with which we will eventually 
communicate sending signals at $t$. The instantaneous 
(at cosmic time $t$) radius of the `event horizon' is
\begin{equation}
d_{e}(t) = a(t) \int ^{\infty}_{t} 
\frac{dt^{\prime}}{a(t^{\prime})}~\cdot
\label{eq:event}
\end{equation}
It is obvious, from this formula, that the `event horizon' is infinite 
for matter or radiation dominated universe. For inflation, however, we 
obtain a slowly varying event horizon with radius
$d_{e}(t) = H^{-1} < \infty$. Points, in our `event horizon' at $t$, 
with which we can communicate sending signals at $t$, are eventually 
pulled away by the `exponential' expansion and we cease to be able to 
communicate with them again emitting signals at later times. We say that 
these points (and the corresponding scales) crossed outside the event 
horizon. The situation is very similar to that of a black hole. Indeed, 
the exponentially expanding (de Sitter) space is like a black hole turned 
inside out. This means that we are inside and the black hole surrounds us 
from all sides. Then, exactly as in a 
black hole, there are quantum fluctuations of the `thermal type' governed 
by the `Hawking temperature'~\cite{hawking,gibbons} 
$T_{H} = H/2\pi$.
It turns out~\cite{bunch,vilenkin} that the quantum fluctuations of all 
massless fields (the inflaton is nearly massless due to the `flatness' 
of the potential) are $\delta \phi = H/ 2\pi = T_{H}$. These 
fluctuations of $\phi$ lead to energy density fluctuations
$\delta \rho = V^{\prime} (\phi) \delta \phi$. As the scale of 
this perturbations crosses outside the event horizon, they 
become~\cite{fischler} classical metric perturbations.

\par
The evolution of these fluctuations outside the `inflationary horizon' 
is quite subtle and involved due to the gauge freedom
in general relativity. However, there is a simple gauge invariant 
quantity~\cite{zeta} $\zeta \approx \delta \rho/(\rho +p)$, 
which remains constant outside the horizon. Thus, the density fluctuation
at any present physical (comoving) scale 
$\ell,~(\delta \rho/\rho)_{\ell}$, when this 
scale crosses inside the `post-inflationary' particle horizon ($p$=0 at 
this instance) can be related to the value of $\zeta$ when the same scale 
crossed outside the inflationary event horizon 
(symbolically at $\ell \sim H^{-1}$). 
This latter value of $\zeta$ can be found using Eq.\ref{eq:infeq} 
and turns out to be
\begin{equation}
\zeta \mid_{\ell \sim H^{-1}} = 
\left(\frac {\delta \rho}{\dot {\phi}^2}\right)
_{\ell \sim H^{-1}} =
\left(\frac {V^{\prime} (\phi) H(\phi)}
{2 \pi \dot{\phi}^2}\right)_{\ell \sim H^{-1}}=
- \left(\frac {9 H^{3}(\phi)}
{2 \pi V^{\prime}(\phi)}\right)_{\ell \sim H{-1}}~.
\label{eq:zeta}
\end{equation}
Taking into account an extra 2/5 factor from the fact that the 
universe is matter dominated when the scale $\ell$ re-enters the 
horizon, we obtain
\begin{equation}
\left(\frac {\delta \rho}{\rho}\right)_{\ell} = 
\frac {16 \sqrt {6 \pi}}{5}~\frac
{V^{3/2}(\phi_{\ell})}{M^{3}_{P} V^{\prime}
(\phi_{\ell})} ~\cdot
\label{eq:deltarho}
\end{equation}

\par
The calculation of $\phi_{\ell}$, the value of the inflaton field 
when the `comoving' scale $\ell$ crossed outside the event horizon, 
goes as follows. A `comoving' (present physical) scale $\ell$, 
at $T_r$, was equal to $\ell (a(t_{d})/a(t_{0})) = 
\ell(T_{0}/T_{r})$.  Its magnitude at the end of inflation 
($t=t_{f}$) was $\ell (T_{0}/T_{r})(a(t_{f})/a(t_{d})) = 
\ell (T_{0}/T_{r})(t_{f}/t_{d})^{(\nu +2)/3 \nu}
\equiv \ell_{{\rm{phys}}}(t_{f})$, where the potential 
$V(\phi)=\lambda\phi^{\nu}$ was assumed. The scale $\ell$, 
when it crossed outside the inflationary horizon, was equal to
$H^{-1}(\phi_{\ell})$. We, thus, obtain
\begin{equation}
H^{-1}(\phi_{\ell}) e^{N(\phi_{\ell})} = 
\ell_{{\rm{phys}}}(t_{f})~.
\label{eq:lphys}
\end{equation}
Solving this equation, one can calculate $\phi_{\ell}$ and, 
subsequently, $N(\phi_{\ell})\equiv N_{\ell}$, the number of 
e-foldings the scale $\ell$ suffered during inflation. In particular, 
for our present horizon scale $\ell \approx 2H_{0}^{-1} \sim
10^4$~Mpc ($H_{0}$ is the present value of the Hubble parameter), 
it turns out that $N_{H_{0}}\approx 50-60$.

\par
Now, taking the potential $V(\phi)= \lambda \phi^4$,
~Eqs.\ref{eq:expefold}, \ref{eq:deltarho} and \ref{eq:lphys} give
\begin{equation}
\left(\frac {\delta \rho} {\rho}\right)_{\ell} = 
\frac {4 \sqrt{6 \pi}} {5}
\lambda^{1/2}\left(\frac{\phi_{\ell}}{M_{P}}\right)^3 = 
\frac {4 \sqrt{6 \pi}}{5}
\lambda^{1/2} \left(\frac {N_{\ell}}{\pi}\right)^{3/2}~.
\label{eq:nl}
\end{equation}
The measurements of COBE~\cite{cobe}, $ (\delta \rho/
\rho)_{H_{0}} \approx 6 \times 10^{-5}$, then imply that 
$\lambda \approx 6 \times 10^{-14}$ for $N_{H_{0}} \approx 55$. 
Thus, we see that the inflaton must be a very weakly coupled field.
In non-supersymmetric GUTs, the inflaton must necessarily be a gauge 
singlet field since otherwise radiative corrections will certainly make 
it strongly coupled. This is, undoubtedly, not a very satisfactory situation 
since we are forced to introduce an otherwise unmotivated extra 
{\it ad hoc} very weakly coupled gauge singlet.
In supersymmetric GUTs, however, the inflaton could be identified
~\cite{nonsinglet} with a conjugate pair of gauge non-singlet fields 
$\phi$, $\bar{\phi}$, already existing in the theory and causing 
the gauge symmetry breaking. Absence of strong radiative corrections 
from gauge interactions is guaranteed, in this case, by the mutual 
cancellation of the D-terms of these fields.

\par
The spectrum of the density fluctuations emerging from inflation can also
be analyzed. For definiteness, we will again take the potential
$V(\phi) =\lambda \phi^{\nu}$. One then finds that 
$(\delta \rho/ \rho)_{\ell}$ is proportional to 
$\phi_{\ell}^{(\nu+2)/2}$ which, combined with the fact that
$N(\phi_{\ell})$ is proportional to $\phi_{\ell}^{2}$ 
(see Eq.\ref{eq:expefold}), gives
\begin{equation}
\left(\frac {\delta \rho}{\rho}\right)_{\ell} = 
\left(\frac{\delta \rho}{\rho}\right)_{H_{0}}
\left(\frac{N_{\ell}}{N_{H_{0}}}\right)^{\frac{\nu+2}{4}}.
\label{eq:spectrum}
\end{equation}
The scale $\ell$ divided by the size of our present horizon  
($\approx 10^4~{\rm{Mpc}}$) should equal exp$(N_{\ell} - 
N_{H_{0}})$. This gives $N_{\ell}/N_{H_{0}} =
1 + \ell n(\ell/10^{4})^{1/N_{H_{0}}}$ which expanded around
$\ell \approx 10^4$~Mpc and substituted in Eq.\ref{eq:spectrum} 
yields
\begin{equation}
\left(\frac{\delta \rho}{\rho}\right)_{\ell} = 
\left(\frac{\delta \rho}{\rho}\right)_{H_{0}}
\left(\frac {\ell}{10^4~{\rm{Mpc}}}\right)^{\alpha_{s}}~,
\label{eq:alphas}
\end{equation}
with $\alpha_{s}=(\nu + 2) /4 N_{H_{0}}$. For $\nu=4$, 
$\alpha_{s} \approx 0.03$ and the fluctuations are essentially 
scale independent.

\section{Density Fluctuations in `Matter'}\label{sec:matter}

We will now discuss the evolution of the primordial density fluctuations
after their scale enters the post-inflationary horizon. To this end, 
we introduce~\cite{bardeen} the `conformal' time, $\eta$, so that the 
Robertson-Walker metric takes the form of a conformally expanding 
Minkowski space:
\begin{equation}
ds^2 = -dt^2 + a^2 (t)~d\bar{r}^2 = a^{2}(\eta)~ 
(-d \eta^2 + d\bar{r}^2)~,
\label{eq:conf}
\end{equation}
where $\bar{r}$ is a comoving 3-vector.
The Hubble parameter now takes the form $H=\dot{a}(t)/a(t) = 
a^{\prime}(\eta)/a^{2}(\eta)$
and the Friedmann equation can be rewritten as
\begin{equation}
\frac{1}{a^{2}}\left(\frac{a^{\prime}}{a}\right)^{2} = 
\frac {8 \pi G}{3} \rho~,
\label{eq:conffried}
\end{equation}
where primes denote derivation with respect to the `conformal' time 
$\eta$. The continuity equation takes the form $ \rho^{\prime} = 
- 3 \tilde {H} (\rho+p)$ with $\tilde{H} = a^{\prime}/a$. 
For a matter dominated universe, $\rho \propto a^{-3}$
which gives $a=(\eta/\eta_{0})^2$ and $a^{\prime}/a =
2/\eta$ ($\eta_0$ is the present value of $\eta$). 

\par
The `Newtonian' Eq.\ref{eq:newton} can now be written in the form
\begin{equation}
\delta^{\prime \prime}_{\bar{k}} (\eta) + 
\frac {a^{\prime}}{a} \delta^{\prime}_{\bar{k}}(\eta) 
- 4 \pi G \rho a^{2} \delta_{\bar{k}} (\eta) =0~,
\label{eq:confnewton}
\end{equation}
and the growing (Jeans unstable) mode $\delta_{\bar{k}} (\eta)$ 
is proportional to $\eta^{2}$ and can be expressed~\cite{schaefer} as
\begin{equation}
\delta_{\bar{k}} (\eta) = \epsilon_{H} 
\left(\frac {k \eta}{2}\right)^{2} \hat {s} (\bar{k})~,
\label{eq:growmode}
\end{equation}
where $\hat{s}(\bar{k})$ is a Gaussian random variable satisfying
\begin{equation}
<\hat{s}(\bar {k})> = 0~,~<\hat{s}(\bar {k}) 
\hat {s}(\bar {k}^{\prime})>
= \frac {1} {k^{3}} \delta (\bar{k} - \bar {k}^{\prime})~,
\label{eq:gauss}
\end{equation}
and $\epsilon_{H}$ is the amplitude of the perturbation when its 
scale crosses inside the `post-inflationary horizon' . The latter 
can be seen as follows. A `comoving' (present physical) length $\ell$
crosses inside the post-inflationary horizon when  $a \ell/ 2 \pi = 
H^{-1}=a^2/a^{\prime}$ which gives $\ell/ 2\pi \equiv k^{-1} = 
a/a^{\prime} = \eta_{H}/2$ or $k \eta_{H}/2 = 1$, where 
$\eta_{H}$ is the conformal time at horizon crossing.
This means that, at horizon crossing, $\delta_{\bar{k}} (\eta_{H}) = 
\epsilon_{H}\hat{s} (\bar{k})$.
For scale invariant perturbations, the amplitude $\epsilon _{H}$ is
constant. The gauge invariant perturbations of the `scalar gravitational 
potential' are given~\cite{bardeen} by the Poisson's equation,
\begin{equation}
\Phi = - 4 \pi G \frac {a^2}{k^2}\rho \delta_{\bar{k}}(\eta)~.
\label{eq:poisson}
\end{equation}
From the  Friedmann Eq.\ref{eq:conffried}, we then obtain
\begin{equation}
\Phi = - \frac {3}{2} \epsilon_{H} \hat{s} (\bar{k})~.
\label{eq:scalarpot}
\end{equation}

\par
The spectrum of the density perturbations can be characterized by the 
correlation function
\begin{equation}
\xi(\bar{r}) \equiv <\tilde{\delta}^{*}(\bar{x},\eta)
\tilde{\delta}(\bar{x} + \bar{r},
\eta)>~,
\label{eq:corr}
\end{equation}
where
\begin{equation}
\tilde{\delta}(\bar{x},\eta) = \int d^{3}k 
\delta_{\bar{k}}(\eta)e^{i\bar{k}\bar{x}}~.
\label{eq:fourier}
\end{equation}
Substituting Eq.\ref{eq:growmode} in Eq.\ref{eq:corr} and using 
Eq.\ref{eq:gauss}, we obtain
\begin{equation}
\xi (\bar{r}) = \int d^{3}k e^{-i\bar{k}\bar{r}} \epsilon^2_{H}
\left(\frac {k \eta}{2}\right)^{4} \frac{1}{k^{3}}~,
\label{eq:index}
\end{equation}
and, thus, the spectral function  $P(k, \eta) = 
\epsilon^2_{H}(\eta^{4}/16)k$ is proportional to $k$ for 
$\epsilon_{H}$ constant. We say that, in this case, the
`spectral index' $n=1$ and we have a Harrison-Zeldovich~\cite{hz} 
flat spectrum. In the general case, $P \propto k^{n}$ with $n = 
1-2 \alpha_{s}$ (see Eq.\ref{eq:alphas}). For $V(\phi)= 
\lambda \phi^{4}$, we get $n \approx 0.94$.

\section{Temperature Fluctuations} \label{sec:temperature}

The density inhomogeneities produce temperature fluctuations in the CBR.
For angles $\theta \stackrel{_{>}}{_{\sim }} 2~^o$, the dominant 
effect is the scalar Sachs-Wolfe~\cite{sachswolfe} effect. Density 
perturbations on the `last scattering surface' cause scalar gravitational 
potential fluctuations, $\Phi$, which, in turn, produce temperature 
fluctuations in the CBR. The physical reason is that regions with a
deep gravitational potential will cause the photons to lose energy as 
they climb up the well and, thus, appear cooler. For 
$\theta \stackrel{_{<}}{_{\sim }} 2~^o$, the dominant effects 
are: i) Motion of the last scattering surface causing Doppler shifts, 
and ii) Intrinsic fluctuations of the photon temperature, $T_{\gamma}$, 
which are more difficult to calculate since they depend on microphysics,
the ionization history, photon streaming and other effects.

\par
The temperature fluctuations at an angle $\theta$ due to the scalar 
Sachs-Wolfe effect turn out~\cite{sachswolfe} to be given by 
$(\delta T/T)_{\theta} = - \Phi_{\ell}/3$, where $\ell$ is the
`comoving' scale on the `last scattering surface' which subtends the angle
$\theta$ [~$\ell \approx 100~h^{-1}(\theta/{\rm{degrees}})
~{\rm{Mpc}}~]$ and $\Phi_{\ell}$ the
corresponding scalar gravitational potential fluctuations. 
From Eq.\ref{eq:scalarpot}, we then obtain $(\delta T/T)_{\theta} = 
(\epsilon_{H}/2) \hat{s} (\bar{k})$, which using 
Eq.\ref{eq:growmode} becomes
\begin{equation}
\left(\frac{\delta T}{T}\right)_{\theta} = 
\frac{1}{2} \delta_{\bar{k}}(\eta_{H})
= \frac {1}{2} \left(\frac{\delta \rho}
{\rho}\right)_{\ell \sim 2\pi k^{-1}}~.
\label{eq:swe}
\end{equation}
The COBE scale (our present horizon )corresponds to $\theta 
\approx 60~^o$. Eqs.\ref{eq:expefold}, \ref{eq:deltarho} and
\ref{eq:swe} give
\begin{equation}
\left(\frac {\delta T} {T}\right)_{\ell} 
\propto \left(\frac{\delta \rho}{\rho}\right)_{\ell} 
\propto \frac{V^{3/2}(\phi_{\ell})}{M^{3}_{P} V^{\prime}
(\phi_{\ell})} \propto N_{\ell}^{\frac{\nu+2}{4}}~.
\label{ewq:tempflu}
\end{equation}
Analyzing the temperature fluctuations in spherical harmonics we can 
obtain the quadrupole anisotropy due to the scalar Sachs-Wolfe effect,
\begin{equation}
\left(\frac{\delta T}{T}\right)_{Q-S} = \left(\frac{32 \pi}
{45}\right)^{1/2} \frac{V^{3/2}(\phi_{\ell})}{M^{3}_{P}
V^{\prime}(\phi_{\ell})}~\cdot
\label{eq:quadrupole}
\end{equation}
For $V(\phi) = \lambda \phi^{\nu}$, this becomes
\begin{equation}
\left(\frac{\delta T}{T}\right)_{Q-S} = 
\left(\frac{32 \pi}{45}\right)^{1/2} 
\frac{\lambda^{1/2}\phi_{\ell}^{\frac{\nu+2}{2}}}
{\nu M^{3}_{P}}= \left(\frac{32 \pi}{45}\right)^{1/2}
\frac {\lambda^{1/2}}{\nu M^{3}_{P}}\left(\frac{\nu M^{2}_{P}}
{4 \pi}\right)^{\frac{\nu+2}{4}} N_{\ell}^{\frac{\nu+2}{4}}~.
\label{eq:anisotropy}
\end{equation}
Comparing this with the COBE~\cite{cobe} measurements, 
$(\delta T/T)_{Q} \approx 6.6 \times 10^{-6}$, we obtain 
$\lambda \approx 6 \times 10^{-14}$, for $\nu=4$
and number of e-foldings suffered by our present horizon scale during
inflation $N_{\ell \sim H^{-1}_{0}} \equiv N_{Q} \approx 55$.

\par
There are also `tensor'~\cite{tensor} (gravitational wave) fluctuations 
in the temperature of CBR. The quadrupole tensor anisotropy is
\begin{equation}
\left(\frac{\delta T}{T}\right)_{Q-T} \approx 0.77~\frac 
{V^{1/2}(\phi_{\ell})}{M^{2}_{P}}~\cdot
\label{eq:tensor}
\end{equation}
The total quadrupole anisotropy is given by
\begin{equation}
\left(\frac{\delta T}{T}\right)_{Q} = 
\left[ \left(\frac{\delta T}{T}\right)
^{2}_{Q-S} + \left(\frac{\delta T}{T}\right)^{2}
_{Q-T}\right]^{1/2}~,
\label{eq:total}
\end{equation}
and the ratio
\begin{equation}
r = \frac{\left(\delta T/T\right)^{2}_{Q-T}}
{\left(\delta T/T\right)^{2}_{Q-S}} \approx 0.27
~\left(\frac{M_{P} V^{\prime}(\phi_{\ell})}
{V(\phi_{\ell})}\right)^{2}~\cdot
\label{eq:ratio}
\end{equation}
For $V(\phi) = \lambda \phi^{\nu}$, we obtain $r 
\approx 3.4~\nu/N_{H}\ll 1$, and the `tensor' contribution to the 
temperature fluctuations of the CBR is negligible.

\section{Hybrid Inflation} \label{sec:hybrid}
\subsection{The non Supersymmetric Version} \label{subsec:nonsusy}

The most important disadvantage of the inflationary scenarios described 
so far is that they need extremely small coupling constants in order 
to reproduce the results of COBE~\cite{cobe}. This difficulty was 
overcome some years ago by Linde~\cite{hybrid} 
who proposed, in the context of non-supersymmetric GUTs, 
a clever inflationary  scenario known as hybrid inflation. 
The  idea was to use two real scalar  fields $\chi$ and $\sigma$ 
instead of one that was normally used. The  field $\chi$
provides the vacuum energy which drives inflation  while  $\sigma$ is 
the slowly  varying  field during inflation. The main advantage of this 
scenario is that it can reproduce the observed temperature fluctuations 
of the CBR  with `natural'  values  of the parameters in contrast to 
previous realizations of inflation (like the `new'~\cite{new}
or `chaotic'~\cite{chaotic} inflationary scenarios). The potential 
utilized by Linde is
\begin{equation}
V ( \chi, \sigma)= \kappa^2 \left( M^2 - 
\frac {\chi^2}{4}\right)^2 +   \frac{\lambda^2 
\chi^2 \sigma^2}   {4}   +   \frac {m^2\sigma^2}{2}~~,
\label{eq:lindepot}
\end{equation}
where  $\kappa,~\lambda$ are dimensionless positive  coupling 
constants and $M$, $m$ mass parameters. The vacua lie  at $\langle 
\chi\rangle=  \pm 2 M$, $\langle \sigma \rangle=0$. Putting 
$m$=0, for  the moment, we observe that the potential 
possesses an  exactly flat direction  at  $\chi=0$ 
with $V(\chi=0 ,\sigma)=\kappa^2 M^4$. The mass squared of the 
field $\chi$ along  this flat  direction is given by $m^2_\chi = 
- \kappa^2 M^2 +  \frac{1}{2}  \lambda^2  \sigma^2$  and  remains 
non-negative  for  $\sigma \geq \sigma_c = \sqrt  {2}  \kappa M/  
\lambda $. This means that, at $\chi=0$ and $\sigma \geq  \sigma_c$, 
we obtain a valley of minima with  flat bottom. Reintroducing  the mass 
parameter $m$ in Eq.\ref{eq:lindepot}, we observe that  
this  valley acquires a non-zero  slope. A region of the universe, 
where $\chi$ and $\sigma$  happen to be almost uniform with 
negligible kinetic  energies and  with  values close to the bottom of 
the  valley  of minima, follows  this valley in its subsequent  evolution 
and undergoes  inflation. The quadrupole anisotropy of CBR produced during 
this inflation can be estimated, from Eq.\ref{eq:quadrupole}, to be
\begin{equation}
\left(\frac {\delta T}{T}\right)_{Q} 
\approx  \left(\frac {16 \pi}{45}\right)^{1/2} 
\frac{\lambda \kappa^2 M^5}
{M^3_Pm^2}~.
\label{eq:lindetemp}
\end{equation}
The COBE~\cite{cobe} result, 
$(\delta T/T)_{Q} \approx 6.6 \times 10^{-6}$, can then be reproduced 
with $M \approx 2.86 \times 10^{16}$ GeV (the supersymmetric GUT vev) 
and $m \approx 1.3~\kappa \sqrt {\lambda}\times 10^{15}$ GeV 
$ \sim 10^{12}$ GeV for $\kappa, \lambda \sim 10^{-2}$.
Inflation terminates abruptly at $\sigma=\sigma_{c}$ and is followed by
a `waterfall', i.e., a sudden entrance into  an oscillatory phase about a 
global minimum. Since the system can fall into either of the two available 
global minima with equal probability, topological defects are copiously 
produced if they are predicted by the particular particle physics model 
one is considering.

\subsection{The Supersymmetric Version}\label{subsec:susy}

The hybrid inflationary scenario is~\cite{lyth} `tailor made' for 
application to supersymmetric GUTs except that the mass of $\sigma$, 
$m$, is unacceptably large for supersymmetry, where all scalar fields 
acquire masses of order $m_S \sim 1$ TeV from soft supersymmetry 
breaking. To see this, consider a supersymmetric GUT with a (semi-simple) 
gauge group $G$ of rank $\geq 5$ with $G \to G_S$ 
(the Standard Model gauge group) at a scale 
$M \sim 10^{16}$GeV. The spectrum of the theory below $M$ is assumed 
to coincide with the Minimal Supersymmetric Standard Model (MSSM) spectrum  
plus Standard Model singlets so that the successful predictions for 
$\alpha_{s}$, ${\rm{sin}}^{2} \theta_{W}$ are retained. The theory 
may also possess global symmetries. The breaking of $G$ is
achieved through the superpotential
\begin {equation}
W = \kappa S (- M^2 + \bar{\phi} \phi),
\label{eq:superpot}
\end {equation}
where $\bar{\phi}, \phi$ is a conjugate pair of Standard Model singlet 
left handed superfields which belong to non-trivial representations of 
$G$ and reduce its rank by their vevs and $S$ is a gauge singlet left 
handed superfield. The coupling constant $\kappa$ and the mass parameter 
$M$ can be made real and positive by suitable redefinitions of the phases
of the superfields. This superpotential is the most general renormalizable 
superpotential consistent with a $U(1)$ $R$ symmetry under which 
$W \to e^{i\theta} W,~S \to e^{i \theta}S,
~\bar {\phi} \phi \to \bar{\phi} \phi$ and gives the potential
\begin{eqnarray*}
V=\kappa^2 \mid M^2 - \bar{\phi} \phi \mid^2 
 +\kappa^2 \mid S \mid^2
(\mid \phi \mid^2  + \mid \bar{\phi}\mid^2)
\end{eqnarray*}
\begin{equation}
+{\rm{ D-terms}}.
\label{eq:hybpot}
\end{equation}
Restricting ourselves to the D-flat direction $\bar{\phi}^* = 
\phi$ which contains the supersymmetric  minima and performing 
appropriate gauge and $R$ transformations, we can bring $S$, 
$\bar{\phi}$, $\phi$ on the real axis, i.e., $S \equiv \sigma/
\sqrt{2}$, $\bar{\phi}=\phi \equiv \chi/2$, where $\sigma$, 
$\chi$ are normalized real scalar fields. The potential then takes 
the form in Eq.\ref{eq:lindepot} with $\kappa = \lambda$
and $m=0$ and, thus, Linde's potential for hybrid inflation is almost
obtainable from supersymmetric GUTs but without the mass term of 
$\sigma$ which is, however, of crucial importance since it provides 
the slope of the valley of minima necessary for inflation.

\par
One way to obtain a valley of minima useful for inflation is~\cite{lp} 
to replace the renormalizable trilinear term in the superpotential $W$ 
in Eq.\ref{eq:superpot} by the next order non-renormalizable coupling. 
Another way, which we will adopt here, is~\cite{dss} to keep the 
renormalizable superpotential in Eq.\ref{eq:superpot} and use the 
radiative corrections along the inflationary valley
($\phi = \bar{\phi} = 0$~, $S > S_{c} \equiv M$).
In fact, due to supersymmetry breaking by the `vacuum' 
energy density ,~$\kappa^{2} M^{4}$, along this valley, there 
are important radiative corrections. At one loop, and for $S$ 
sufficiently larger than $S_{c}$, the inflationary potential is 
given~\cite{dss,lss} by
\begin{equation}
V_{{\rm{eff}}} (S) = \kappa^2 M^4 
\left[ 1 + \frac{\kappa^2}{16\pi^2} \left( \ell n
\left(\frac{\kappa^{2} S^{2}}{\Lambda^2}\right) + 
\frac{3}{2} - \frac{S_c^4}{12S^4} + \cdots \right)\right]~,
\label{eq:veff}
\end{equation}
where $\Lambda$ is a suitable mass renormalization scale. 
Using this effective potential and Eq.\ref{eq:quadrupole}, one finds 
that the cosmic microwave quadrupole anisotropy 
$(\delta T/T)_{Q} \approx 8 \pi (N_{Q}/45)^{1/2} (x_{Q}/y_{Q}) 
(M/M_{P})^2$. Here $y_{Q}=
x_{Q}(1-7/(12x_{Q}^2)+\cdots)$ with $x_{Q} = S_{Q}/M$, and 
$S_{Q}$ is the value of the scalar field $S$ when the scale which 
evolved to the present horizon size crossed outside the de Sitter 
(inflationary) horizon. Also from Eq.\ref{eq:veff}, one finds 
$\kappa \approx (8\pi^{3/2}/\sqrt{N_{Q}})~ y_{Q}~(M/M_{P})$.  

The inflationary phase ends as $S$ approaches $S_{c}$ from above. 
Writing $S=xS_{c}$, $x=1$ corresponds to the phase transition from 
$G$ to $G_{S}$ which, as it turns out, more or less coincides 
with the end of the inflationary phase (this is checked by noting the 
amplitude of the quantities $\epsilon$ and $\eta$ in Eq.\ref{eq:src}). 
Indeed, the $50-60$ e-foldings needed for the inflationary scenario can 
be realized even with small values of $x_{Q}$. For definiteness, we 
will take $x_{Q}\approx 2$ (see Sec.\ref{sec:neutrinos}). From 
COBE~\cite{cobe} one then obtains 
$M \approx 5.5 \times 10^{15} $GeV and $\kappa \approx 4.5 
\times 10^{-3}$ for $N_{Q} \approx 56$. Moreover, the primordial 
density fluctuation spectral index $n \simeq 0.98$. We see that the 
relevant part of inflation takes place at $S \sim 10^{16}$ GeV. 
An important consequence of this is~\cite{lyth,lss,sugra} that the 
supergravity corrections can be negligible.

In conclusion, it is important to note that the superpotential $W$ in 
Eq.\ref{eq:superpot} leads to hybrid inflation in a `natural' way.
This means that a) there is no need of very small coupling constants, 
b) $W$ is the most general renormalizable superpotential allowed by 
the gauge and  $R$ symmetries, and c) supersymmetry guarantees that 
the radiative corrections do not invalidate inflation. They rather 
provide a slope along the inflationary trajectory which drives the 
inflaton towards the supersymmetric vacua. 

\section{`Reheating' in Supersymmetric Hybrid Inflation}
\label{sec:reheathybrid}

In order to discuss the `reheating' of the universe after inflation, 
we need to be a little more specific about the underlying particle 
physics model. To this end, we consider~\cite{lss} a supersymmetric 
model based on the left-right symmetric gauge group $G_{LR} = 
SU(3)_{c} \times SU(2)_{L} \times SU(2)_{R} \times U(1)_{B-L}$.
Of course, it is anticipated that $G_{LR}$ is embedded in a grand 
unified theory such as $SO(10)$ or $SU(3)_c \times SU(3)_L 
\times SU(3)_R$. The breaking of $G_{LR}$ to $G_{S}$ is achieved 
by the renormalizable superpotential in Eq.\ref{eq:superpot},
where $\phi, \bar{\phi}$ are now identified with the
Standard Model singlet components of a conjugate pair of $SU(2)_R 
\times U(1)_{B-L}$ doublet left handed superfields.

\par
The inflaton (oscillating system) consists of the two complex scalar 
fields $S$ and $\theta=(\delta \phi + \delta
\bar{\phi})/\sqrt{2}$, where $\delta \phi = \phi - M$, 
$\delta \bar{\phi} = \bar{\phi} - M$, with
mass $m_{infl} = \sqrt{2}\kappa M$. We mainly concentrate on the 
decay of $\theta$. Its relevant coupling to `matter' is 
provided~\cite{lss} by the non-renormalizable superpotential coupling 
(in symbolic form) 
\begin{equation}
\frac{1}{2}\left( \frac{M_{\nu^c}}{M^2} \right) 
\bar{\phi} \bar{\phi} \nu^c \nu^c~,
\label{eq:coupling}
\end{equation}
where $M_{\nu^c}$ denotes the Majorana mass of the relevant right 
handed neutrino $\nu^c$. Without loss of generality, we assume that 
the Majorana mass matrix of the right handed neutrinos has been 
brought to diagonal form with positive entries. Clearly,
$\theta$ decays predominantly into the heaviest right handed neutrino
permitted by phase space, i.e., with $M_{\nu^{c}} \leq m_{infl}/2$. 

\par
The field $S$ is not important for reheating since it 
can rapidly decay~\cite{lss} into higgsinos through 
the renormalizable superpotential term $\xi S h^{(1)} h^{(2)}$ 
allowed by gauge symmetry, where $h^{(1)},\ h^{(2)}$ denote the 
electroweak higgs doublets which couple to the up and down type
quarks respectively, and $\xi$ is a suitable coupling constant. 
Note that, after supersymmetry breaking, $S$ acquires~\cite{lss,dvali} 
an expectation value $\langle S\rangle\sim m_S$ ($m_S\sim$ 1 TeV 
being the magnitude of supersymmetry breaking in the visible sector) 
and generates the $\mu$ term.

\par
Following standard procedures (see Eq.~\ref{eq:reheat}), and assuming 
the MSSM spectrum, the `reheating' temperature $T_r$ is found to be 
given by
\begin{equation}
T_r\ \approx \frac{1}{7} \left( \Gamma_\theta M_P\right)^{1/2}, 
\label{eq:reheat1}
\end{equation}
where $\Gamma_\theta \approx (1/ 16\pi) (\sqrt{2} M_{\nu^c}/M)^2 
\sqrt{2}\kappa M$ is the decay rate of $\theta$. Substituting $\kappa$ 
as a function of $N_Q$, $y_Q$ and $M$, we find~\cite{lss}
\begin{equation}
T_r \approx \ {1\over 12} \left( \frac{56}{N_Q} \right)^{1/4} 
\sqrt{y_Q}\ M_{\nu^c}.
\label{eq:reheat2}
\end{equation}
For $x_{Q} \approx $ 2 and $N_{Q} \approx 56$,
we have $T_{r} \approx M_{\nu^{c}}/ 9.23$. We will assume that 
$T_{r}$ is restricted by the gravitino constraint~\cite{gravitino}, 
$T_{r}\stackrel{_<}{_\sim} 10 ^9 $ GeV. Note that $T_{r}$ is 
closely linked to the mass of the heaviest $\nu^{c}$ satisfying 
$M_{\nu^{c}} \leq m_{infl}/2$.

\section{Baryogenesis via Leptogenesis} \label{sec:baryons}
\subsection{Primordial Leptogenesis} \label{subsec:leptons}

In the hybrid inflationary models under consideration here, it is 
not convenient to produce the baryon asymmetry of the universe (BAU) 
in the usual way, i.e., through the decay of color $3,~\bar{3}$ 
fields $(g,~g^{c})$. Some of the reasons are the following: i) 
For theories where leptons and quarks belong to different 
representations of the unifying gauge group $G$ (which is the case, 
for example, for $G=G_{LR}$ or $SU(3)_c \times SU(3)_L 
\times SU(3)_R$), the baryon number can be made almost exactly 
conserved by imposing an appropriate discrete symmetry. In particular, 
for $G=G_{LR}$, we can impose~\cite{baryonparity} a discrete 
symmetry under which $q\rightarrow - q$,~$q^{c} \rightarrow 
- q^{c}$,~$\bar{q} \rightarrow - \bar{q}$,~$\bar{q}^{c} 
\rightarrow - \bar{q}^{c}$ and all other superfields
remain invariant ($q,q^{c},\bar{q},\bar{q}^{c}$ are 
superfields with the quantum numbers of the quarks, antiquarks 
and their conjugates respectively). ii) For theories where such 
a discrete symmetry is absent, we could, in principle, use as 
inflaton a pair of conjugate Standard Model singlet superfields 
$N$,~$\bar{N}$ which decay into $g$,~$g^{c}$. For $G=SU(3)_c 
\times SU(3)_L \times SU(3)_R$, for example, $N$ ($\bar{N}$) 
could be the Standard Model singlet component of the (1, $\bar{3}$, 3) 
( (1, 3, $\bar{3}$) ) superfields with zero $U(1)_{B-L}$ 
charge. But this is again unacceptable since the breaking of 
$SU(3)_c \times SU(3)_L \times SU(3)_R$ by the vevs of $N$,
~$\bar{N}$ predicts~\cite{trinifiedmonopoles} 
magnetic monopoles which can then be copiously produced after inflation. 
Also $T_{r} \stackrel{_{<}}{_{\sim }} 10^9$ GeV 
(gravitino constraint~\cite{gravitino}) 
implies $m_g \stackrel{_{<}}{_{\sim }} 10^{10}$ GeV 
(~from the coupling ($m_g / \langle N\rangle) Ngg^{c}$~) which leads 
to strong deviation from MSSM and possible proton decay problems.

\par
So it is preferable to produce first a primordial lepton asymmetry
~\cite{fy} which can then be partially turned into the observed 
baryon asymmetry of the universe by the non-perturbative `sphaleron' 
effects~\cite{sphaleron} of the electroweak sector. In the particular 
model based on $G_{LR}$ which we consider here, this is the only way 
to produce the BAU since the inflaton decays into right handed neutrinos. 
Their subsequent decay into ordinary higgs particles (higgsinos) and 
light leptons (sleptons) can produce the primordial lepton asymmetry.
It is important, though, to ensure that this primordial lepton asymmetry 
is not erased by lepton number violating $2 \rightarrow 2$ scatterings  
such as $ l l \rightarrow h^{(1)}\,^{*} h^{(1)}\,^{*}$ or 
$l h^{(1)} \rightarrow \bar{l} h^{(1)}\,^{*}$ ($l$ represents 
a lepton doublet) at all temperatures between $T_{r}$ and
100 GeV. This is automatically satisfied since the lepton asymmetry is
protected~\cite{ibanez} by supersymmetry at temperatures between 
$T_{r}$ and $T \sim 10^{7}$ GeV, and for $T 
\stackrel{_{<}}{_{\sim }} 
10^{7}$ GeV, these $2 \rightarrow 2$ scatterings are~\cite{turner} 
well out of equilibrium. 

\par
The lepton asymmetry produced by the out-of-equilibrium decay 
($M_{\nu^{c}_{i}} \gg T_{r})$  of the right handed neutrinos 
$\nu^{c}_{i}$, which emerged from the inflaton decay, is calculated 
to be~\cite{fy,covi}
\begin{equation}
\frac {n_{L}}{s} \approx -  \frac{3}{16 \pi} 
\frac {T_{r}}{m_{infl}}\sum_{l \neq i}
g(r_{li}) \frac{{\rm{Im}}(U~M^{D}\,^{\prime}
~M^{D}\,^{\prime}\,^{\dagger}~U ^{\dagger})^{2}_{il}}
{|\langle h^{(1)}\rangle|^{2}(U~M^{D}\,^{\prime}
~M^{D}\,^{\prime}\,^{\dagger}~U ^{\dagger})_{ii}}~,
\label{eq:genlept}
\end{equation}
where $n_L$ and $s$ are the lepton number and entropy densities, 
$M^{D}\,^{\prime}$ is the diagonal `Dirac' mass 
matrix, $U$ a unitary transformation so that $UM^{D}\,^{\prime}$ 
is the `Dirac' mass matrix in the basis where the `Majorana' mass 
matrix of $\nu^{c}$~'s is diagonal (see Sec.\ref{sec:neutrinos}) 
and $|\langle h^{(1)} \rangle| \approx 174~{\rm{GeV}}$ for 
large ${\rm tan}\beta$. The function
\begin{equation}
g(r_{li}) = r_{li}~\ell n \left(\frac {1 + r^{2}_{li}}
{r^{2}_{li}}\right)~,~r_{li} = \frac {M_{l}} {M_{i}}~,
\label{eq:gfunction}
\end{equation}
with $g(r) \sim 1/r$ as $r \rightarrow \infty$.
Here we took into account the following prefactors:
i) At `reheating', we have $n_{infl} m_{infl} =
(\pi^{2}/30) g_{*} T^{4}_{r}$ ($n_{infl}$ is the number 
density of inflatons) which together with the relation 
$s = (2 \pi^{2}/45) g_{*}T^{3}_{r}$
implies that $n_{infl}/s = (3/4)(T_{r}/m_{infl})$.
ii) Since each inflaton decays into two $\nu^{c}$~'s, their number 
density $n_{\nu^{c}} = 2 n_{infl}$ which then gives $n_{\nu^{c}}/s
= (3/2)(T_{r}/m_{infl})$. iii) Supersymmetry gives an extra
factor of two.

\subsection {Sphaleron Effects} \label{subsec:sphaleron}

To see how the primordial lepton asymmetry partially turns into the
observed BAU, we must first discuss the non-perturbative baryon ($B$-)
and lepton ($L$-) number violation~\cite{thooft} in the Standard Model. 
Consider the electroweak gauge symmetry $SU(2)_{L} \times U(1)_{Y}$ 
in the limit where the Weinberg angle $\theta_W=0$ and concentrate on 
$SU(2)_{L}$ (inclusion of $\theta_{W} \neq 0$ does not alter 
the conclusions). Also, for the moment, ignore the fermions and higgs 
fields so as to have a pure $SU(2)_{L}$ gauge theory. This theory 
has~\cite{vacuum} infinitely many classical vacua which are 
topologically distinct and are characterized by a `winding number' 
$n \in Z$. In the `temporal gauge' ($A_{0}=0$), the remaining gauge 
freedom consists of time independent transformations and the vacuum 
corresponds to a pure gauge
\begin{equation}
A_{i} = \frac{i}{g}~\partial _{i}g(\bar{x}) g ^{-1}(\bar{x})~,
\label{eq:gauge}
\end{equation}
where $g$ is the $SU(2)_{L}$ gauge coupling constant, $\bar{x}$
belongs to 3-space, $i$ =1,2,3, $g(\bar{x}) \in SU(2)_{L}$,  
and $g(\bar{x}) \rightarrow 1 $ as 
$ \mid\bar{x}\mid \rightarrow \infty$.
Thus, the 3-space compactifies to a sphere $S^{3}$ and $g(\bar{x})$ 
defines a map: $S^{3} \rightarrow SU(2)_{L}$ (with the $SU(2)_{L}$
group being topologically equivalent to $S^{3}$). These maps are 
classified into homotopy classes constituting 
the third homotopy group of $S^{3},~\pi_{3}(S^{3})$, and are 
characterized by a `winding number'
\begin{equation}
n = \int d^{3}x~\epsilon^{ijk}~{\rm{tr}} 
\left(\partial_{i}g(\bar{x}) 
g^{-1}(\bar{x})\partial_{j}g(\bar{x}) g^{-1}(\bar{x})
\partial_{k}g(\bar{x}) g^{-1}(\bar{x})\right).
\label{eq:wind}
\end{equation}
The corresponding vacua are denoted as $\mid n\rangle$, $n\in Z$.

\par
The tunneling amplitude from the vacuum $\mid n_{-}\rangle$ at 
$t=-\infty$ to the vacuum $\mid n_{+}\rangle~$ at $t=+\infty$ is 
given by the functional integral
\begin{equation}
\langle n_{+}\mid n_{-}\rangle = \int(dA)~e^{-S(A)}
\label{eq:path}
\end{equation}
over all gauge field configurations satisfying the appropriate 
boundary conditions at $t=\pm \infty$.
Performing a Wick rotation,~ $x_0 \equiv t \rightarrow -i x_{4}$, 
we can go to Euclidean space-time. Any Euclidean field configuration 
with finite action is characterized by an integer topological number 
known as the Pontryagin number
\begin{equation}
q = \frac {g^{2}}{16\pi^{2}} 
\int d^{4}x~{\rm{tr}}\left(F^{\mu \nu} 
\tilde{F}_{ \mu \nu}\right)~,
\label{eq:pontryagin}
\end{equation}
with $\mu$,$\nu$=1,2,3,4 and $\tilde{F}_{\mu \nu}=
\frac {1}{2}\epsilon_{\mu \nu \lambda \rho}F^{\lambda \rho}$ 
being the  dual field strength. But ${\rm{tr}} (F^{\mu \nu} 
\tilde{F}_{\mu \nu}) = \partial ^{\mu} J_{\mu}$, 
where $J_{\mu}$
is the `Chern-Simons current' given by
\begin{equation}
J_{\mu}=\epsilon_{\mu \nu \alpha \beta}~{\rm{tr}}
\left(A^{\nu}
F^{\alpha \beta}-\frac{2}{3}gA^{\nu}A^{\alpha}A^{\beta}\right).
\label{eq:csc}
\end{equation}
In the `temporal gauge' ($A_0=0$),
\begin{eqnarray*}
q=\frac{g^{2}}{16 \pi^{2}} \int d^{4}x~\partial^{\mu}J_{\mu}=
\frac{g^{2}}{16\pi^{2}}\mathop{\Delta}_{x_{4}=\pm \infty} 
\int d^{3}x~J_{0}   
\end{eqnarray*}
\begin{equation}
=\frac{1}{24\pi^{2}}\mathop{\Delta}_{x_{4}=\pm \infty} 
\int d^{3}x~\epsilon^{ijk}~ 
{\rm{tr}}\left(\partial_{i} g g^{-1}\partial_{j}gg^{-1}
\partial_{k}gg^{-1}\right)=n_{+}-n_{-}~.
\label{eq:interpol}
\end{equation}
This means that Euclidean field configurations interpolating between 
the vacua $\mid n_{+}\rangle,~\mid n_{-}\rangle$ at $x_4 =\pm \infty$ 
have Pontryagin number $q= n_{+}-n_{-}$ and the  path integral in 
Eq.\ref{eq:path} should be performed over all these field configurations.

\par
For a given $q$, there is a lower bound on $S(A)$,
\begin{equation}
S(A) \geq \frac{8 \pi^{2}}{g^{2}}\mid q \mid~,
\label{eq:lbound}
\end{equation}
which is saturated if and only if $F_{\mu \nu}=
\pm \tilde{F} _{\mu \nu}$, i.e, if the 
configuration is self-dual or self-antidual. For $q$=1, the
self-dual classical solution is called instanton~\cite{instanton} 
and is given by (in the `singular' gauge)
\begin{equation}
A_{a \mu}(x)=\frac{2 \rho^{2}}{g(x-z)^{2}}
~\frac{\eta_{a \mu \nu} (x-z)^{\nu}}
{(x-z)^{2} + \rho^{2}}~,
\label{eq:instanton}
\end{equation}
where $\eta_{a \mu \nu}$ ($a$=1,2,3; $\mu$,$\nu$=1,2,3,4) 
are the t' Hooft symbols with $\eta_{aij}=
\epsilon_{aij}$ ($i$,$j$=1,2,3), $\eta_{a4i}=-\delta_{ai}$, 
$\eta_{ai4}=\delta_{ai}$ and $\eta_{a44}=0$. The instanton 
depends on four Euclidean coordinates $z_{\mu}$ (its position) 
and its scale (or size) $\rho$. Two successive
vacua $\mid n\rangle$,~$\mid n+1\rangle$ are separated by a 
potential barrier of height $\propto \rho^{-1}$.
The Euclidean action of the interpolating instanton is always equal 
to $8 \pi^{2}/g^{2}$, but the height of the barrier can be made 
arbitrarily small since the size $\rho$ of the instanton can be taken 
arbitrarily large.

\par
We now reintroduce the fermions into the theory and 
observe~\cite{thooft} that the $B$- and $L$- number currents 
carry anomalies, i.e.,
\begin{equation}
\partial_{\mu} J^{\mu}_{B} = \partial _{\mu} J^{\mu}_{L} = 
- n _{g} \frac {g^{2}}{16 \pi ^{2}}~{\rm{tr}} 
(F_{\mu \nu} \tilde{F}^{\mu \nu})~,
\label{eq:anomaly}
\end{equation}
where $n_{g}$ is the number of generations.
It is then obvious that the tunneling from $\mid n_{-}\rangle$ to  
$\mid n_{+}\rangle$ is accompanied by a change of the
$B$- and $L$- numbers, $\Delta B=\Delta L=- n_{g}q=- n_{g} 
(n_{+}-n_{-})$. Note that i) $\Delta (B-L)=0$, and ii) 
for $q$=1, $\Delta B=\Delta L=-3$ which means that we have the 
annihilation of one lepton per family and one quark per family 
and color (12-point function).

\par
We, finally, reintroduce the Weinberg-Salam higgs doublet $\phi$ with 
its vev given by
\begin{equation}
<\phi> = \frac {v}{\sqrt{2}}~\left(\matrix{0 \cr 1 \cr}\right)
~,~v \approx 246 ~{\rm{GeV}}~.
\label{eq:vev}
\end{equation}
It is then easy to see that the instanton ceases to exist as an exact 
solution. It is replaced by the so called `restricted 
instanton'~\cite{restricted} which is an approximate solution 
for $\rho \ll v^{-1}$. For $\mid x-z\mid \ll \rho$, the gauge 
field configuration of the `restricted instanton' essentially
coincides with that of the instanton and the higgs field is
\begin{equation}
\phi(x) \approx \frac {v}{\sqrt{2}}~\left(\frac{(x-z)^{2}}
{(x-z)^{2} + \rho^{2}}\right)^{1/2} 
\left(\matrix {0 \cr 1 \cr } \right)~~.
\label{eq:restricted}
\end{equation}
For $\mid x-z \mid \gg \rho$, the gauge and higgs fields decay to 
a pure gauge and the vev in Eq.\ref{eq:vev} respectively. The action 
of the `restricted instanton' is 
$S_{ri}=(8 \pi^{2}/g^{2})+\pi^{2} v^{2} \rho^{2}+\cdots$,
which implies that the contribution of big size `restricted instantons' 
to the path integral in Eq.\ref{eq:path} is suppressed. This justifies 
{\it a posteriori} the fact that we restricted ourselves to approximate 
instanton solutions with $\rho \ll v^{-1}$.

\par
The height of the potential barrier between the vacua $\mid n\rangle,
~\mid n+1\rangle$ cannot be now arbitrarily small. This can be 
understood by observing that the static energy of the `restricted 
instanton' at $x_{4}=z_{4}$ ($\lambda$ is the higgs self-coupling),
\begin{equation}
E_{b}(\rho) \approx \frac{3 \pi^{2}}{g^{2}}~\frac{1}{\rho} + 
\frac {3}{8}\pi^{2} v^{2} \rho^{2} +  
\frac {\lambda}{4}\pi^{2} v^{4}\rho^{3}~,
\label{eq:static}
\end{equation}
is minimized for
\begin{equation}
\rho_{{\rm{min}}} = \frac {\sqrt{2}}{gv}\left(\frac {\lambda}
{g^{2}}\right)^{-1/2}\left(\left(\frac {1}{64} + 
\frac {\lambda}{g^{2}}\right)^{1/2} - \frac {1}{8}\right)^{1/2} 
\sim M^{-1}_{W}~,
\label{eq:rhomin}
\end{equation}
and, thus, the minimal height of the potential barrier turns out to be
$E_{{\rm{min}}} \sim M_{W} / \alpha_W$ ( $M_{W}$ is the weak 
scale and $\alpha_{W} = g^{2}/4 \pi$).
The static solution which corresponds to the top (saddle point) of 
this potential barrier is called sphaleron~\cite{sphaleronsol} and is 
given by
\begin{equation}
\bar{A} = v~\frac {f(\xi)}{\xi}~\hat{r} \times \bar{\tau}~,
~\phi = \frac {v}{\sqrt{2}}~h (\xi)~ \hat {r}\cdot  
\bar{\tau} \left (\matrix{0 \cr 1 \cr} \right),
\label{eq:sphaleron}
\end{equation}
where $\xi=2M_{W}r$, $\hat{r}$ is the  radial unit vector and the 
3-vector $ \bar{\tau}$ consists of the Pauli matrices. The functions 
$f(\xi),~h(\xi)$, which can be determined numerically, tend to zero 
as $\xi \rightarrow 0$ and to 1 as $\xi \rightarrow \infty$. 
The mass of the sphaleron is estimated to be
\begin{equation}
E_{{\rm{sph}}} = \frac {2M_{W}}{\alpha_{W}}~k,
~1.5 \leq k \leq 2.7~,~ {\rm{for}} ~0 \leq \lambda \leq \infty~,
\label{eq:sphmass}
\end{equation}
and lies between 10 and 15 TeV.

\par
At zero temperature the tunneling from $\mid n\rangle$ to 
$\mid n+1\rangle$ is utterly suppressed~\cite{thooft} by the factor 
exp$(-8 \pi^{2} /g ^{2})$. At high temperatures, however, thermal 
fluctuations over the potential barrier are frequent and this transition 
can occur~\cite{sphaleron} with an appreciable rate. For
$M_{W} \stackrel{_{<}}{_{\sim }} T 
\stackrel{_{<}}{_{\sim }} T_{c}$ ($T_c$ is the 
critical temperature of the electroweak transition), this rate can be 
calculated~\cite{sphaleron} by expanding around the sphaleron 
(saddle point) solution and turns out to be
\begin{equation}
\Gamma \approx 10^{4}~ n_{g}~ \frac {v (T)^{9}}{T^{8}}
~{\rm{exp}} (-E_{{\rm{sph}}}(T)/T)~.
\label{eq:sphrate}
\end{equation}
Assuming that the electroweak phase transition is a second order one, 
$v(T)$ and $E_{{\rm{sph}}}(T) \propto (1 - T^{2}/T^{2}_{c})
^{1/2}$. One can then show that $\Gamma \gg H$ for temperatures $T$
between $\sim 200$ GeV and $\sim T_{c}$. Furthermore, for 
temperatures above $T_{c}$, where the sphaleron solution ceases 
to exist, it was argued~\cite{sphaleron} that we still have 
$\Gamma \gg H$. The overall 
conclusion is that non-perturbative $B$- and $L$- number violating 
processes are in equilibrium in the universe for  
$ T \stackrel{_{>}}{_{\sim }} 200$ GeV. Remember that $B-L$ is conserved 
by these processes.

\par
Given a primordial $L$- number density, one can 
calculate~\cite{ibanez,turner} the resulting $n_{B}/s$ 
($n_B$ is the $B$- number density). In MSSM, the 
$SU(2)_{L}$- instantons produce the effective operator 
(in symbolic form)
\begin{equation}
O_{2} = (q q q l)^{n_{g}} ( \tilde {h}^{(1)} \tilde {h}^{(2)}) 
\tilde{W}^{4}~,
\label{eq:woperator}
\end{equation}
and the $SU(3)_{c}$- instantons the operator
\begin{equation}
O_3 = ( q q u^{c} d^{c})^{n_{g}} \tilde {g}^{6}~,
\label{eq:coperator}
\end{equation}
where $q$, $l$ are the quark, lepton $SU(2)_{L}$- doublets 
respectively, $u^{c}$, $d^{c}$ the up, down type antiquark 
$SU(2)_{L}$- singlets respectively, $g$, $W$ the
gluons and $W$- bosons and tilde represents their superpartners.
We will assume that these interactions together with the usual MSSM 
interactions are in equilibrium at high temperatures.
The equilibrium number density of ultrarelativistic particles 
$\Delta n \equiv n_{{\rm{part}}} - n_{{\rm{antipart}}}$ is 
given by
\begin{equation}
\frac {\Delta n} {s} = \frac {15 g}{4 \pi^{2} g_{*}}
\left (\frac {\mu}{T}\right) 
\epsilon~,
\label{eq:chemical}
\end{equation}
where g is the number of internal degrees of freedom of the particle 
under consideration, $\mu$ its chemical potential and $\epsilon=2$ 
or 1 for bosons or fermions respectively. Each interaction in equilibrium 
implies that the algebraic sum of the chemical potentials of the 
particles involved is zero. Solving these constraints, we end up with 
only two independent chemical 
potentials, $\mu_{q} $ and $\mu_{\tilde{g}}$, and the baryon and 
lepton asymmetries can be expressed~\cite{ibanez} in terms of them
as follows
$$
\frac {n_{B}}{s} = \frac {30}{4 \pi^{2}g_{*}T}
(6n_{g} \mu_{q} - (4n_{g} - 9) \mu_{\tilde{g}})~,
$$
\begin{equation}
\frac {n_{L}}{s} = - \frac {45}{4 \pi^{2} g_{*}T}
\left(\frac {n_{g}(14 n_{g} +9)}
{1+2 n_{g}} \mu_{q} + \Omega (n_{g}) \mu_{\tilde{g}}\right)~,
\label{eq:bla}
\end{equation}
where $\Omega(n_{g})$ is a known~\cite{ibanez} function.
Now soft supersymmetry breaking couplings come in equilibrium at
$T \stackrel{_{<}}{_{\sim }}10^7$ GeV since their rate 
$\Gamma_{S} \approx m^{2}_{S} /T \stackrel{_{>}}{_{\sim }} H 
\approx 30~T^{2}/M_{P}$ ($m_{S}$ is the soft supersymmetry 
breaking scale). In particular, the non-vanishing gaugino mass implies 
$\mu _{\tilde{g}} =0$ and Eqs.\ref{eq:bla} give~\cite{ibanez}
\begin{equation}
\frac {n_{B}}{s} = \frac {4(1+2n_{g})}{22n_{g}+13}
~\frac {n_{B-L}}{s}~.
\label{eq:bbminl}
\end{equation}
Equating $n_{B-L}/s$ with the primordial $n_{L}/s $, we have
$n_{B}/s = (- 28/79) (n_{L}/s)$, for $n_{g}=3$.

\section{Supersymmetric Hybrid Inflation and Neutrinos}
\label{sec:neutrinos}

The supersymmetric model based on the left-right symmetric 
gauge group $G_{LR}$, which we discussed in 
Sec.\ref{sec:reheathybrid}, provides a framework within which hybrid 
inflation, baryogenesis and neutrino oscillations are~\cite{neu} 
closely linked.This scheme, supplemented by a familiar ansatz for 
the neutrino Dirac masses and mixing of the two heaviest 
families and with the MSW resolution of the solar neutrino puzzle, 
implies~\cite{neu} relatively stringent restrictions on the mass of 
the tau-neutrino, $m_{\nu _{\tau }}$, and the $\mu-\tau$ mixing 
angle, $\theta_{\mu \tau }$. 

\par
In order to obtain information about the light neutrino masses and 
mixing of the two heaviest families, we ignore the first family assuming 
it has small mixings. The relevant `asymptotic' (at the GUT scale 
$M_{X}$) $2 \times 2$ mass matrices are $M^{l}$, the mass matrix of 
charged leptons ($l^{c}$,~$l$),~$M^{D}$, the Dirac mass matrix of 
neutrinos ($\nu^{c}$,~$\nu$), and $M^{R}$, the Majorana mass matrix
of $\nu^{c}$ 's. We shall first diagonalize $M^{l}$,~$M^{D}$:
\begin{equation}
M^{l}\rightarrow M^{l}\,^{\prime }=
\tilde{U}^{l^{c}}M^{l}U^{l}=\left( 
\begin{array}{cc}
m_{\mu } &  \\ 
& m_{\tau }
\end{array}
\right) \ ,\label{eq:lep}
\end{equation}
\begin{equation}
M^{D}\rightarrow M^{D}\,^{\prime }=
\tilde{U}^{\nu ^{c}}M^{D}U^{\nu }=\left( 
\begin{array}{cc}
m_{2}^{D} &  \\ 
& m_{3}^{D}
\end{array}
\right) \ ,\label{eq:neut}
\end{equation}
where the diagonal entries are positive. This gives rise to the `Dirac'
mixing matrix $U^{\nu }\,^{\dagger }U^{l}$ in the leptonic charged 
currents. Using the remaining phase freedom, we can bring this matrix 
to the form 
\begin{equation}
U^{\nu }\,^{\dagger }U^{l}\rightarrow \left( 
\begin{array}{cc}
{\cos \theta }^{D} & {\sin \theta }^{D} \\ 
-{\sin \theta }^{D} & {\cos \theta }^{D}
\end{array}
\right)\ ,\label{eq:dirac}
\end{equation}
where $\theta ^{D}~(0\leq \theta ^{D}\leq \pi /2)$ is the `Dirac' 
(not the physical) mixing angle in the $2-3$ leptonic sector.
In this basis, the Majorana mass matrix can be written as 
$M^{R}=U^{-1} M_{0} \tilde{U}^{-1}$, 
where $M_{0}={\rm{diag}}(M_{2},M_{3})$, with $M_{2},~M_{3}$ 
(both positive) being the two Majorana masses, and $U$ is a unitary 
matrix which can be parametrized as 
\begin{equation}
U=\left( 
\begin{array}{cc}
{\cos }\theta & {\sin }\theta \ e^{-i\delta } \\ 
-{\sin }\theta \,e^{i\delta } & {\cos }\theta
\end{array}
\right) \left( 
\begin{array}{cc}
e^{i\alpha _{2}} &  \\ 
& e^{i\alpha _{3}}
\end{array}
\right) \ ,\label{eq:umatrix}
\end{equation}
with $0\leq \theta \leq \pi /2$ and $0\leq \delta <\pi $. 
The light neutrino mass matrix is 
\begin{equation}
m=-\tilde{M}^{D}\,^{\prime }\ 
\frac{1}{M^{R}}\ M^{D}\,^{\prime } 
=\left( 
\begin{array}{cc}
e^{i\alpha _{2}} &  \\ 
& e^{i\alpha _{3}}
\end{array}
\right) \Psi (\theta ,\delta )\left( 
\begin{array}{cc}
e^{i\alpha _{2}} &  \\ 
& e^{i\alpha _{3}}
\end{array}
\right) \  ,\label{eq:mass}
\end{equation}
where $\Psi (\theta, \delta )$ depends also on 
$M_{2}$, $M_{3}$, $m_{2}^{D}$, $m_{3}^{D}$.

We will denote the two positive eigenvalues of the light neutrino mass
matrix by $m_{2}$ (or $m_{\nu _{\mu }}$), $m_{3}$ 
(or $m_{\nu _{\tau }}$). Recall that all the 
quantities here (masses, mixings) are 
`asymptotic'. The determinant and the trace invariance of 
$m^{\dagger }m $ provide us with two constraints on the 
(asymptotic) parameters: 
\begin{equation}
m_{2}m_{3}\ =\ \frac{\left( m_{2}^{D}m_{3}^{D}\right) 
^{2}}{M_{2}\ M_{3}}\ 
,\label{eq:det}
\end{equation}
\begin{equation}
m_{2}\,^{2}+m_{3}\,^{2}\ =\ \frac{\left( m_{2}^{D}\,\,^{2}{\rm c}
^{2}+m_{3}^{D}\,^{2}{\rm s}^{2}\right) ^{2}}{M_{2}\,^{2}}+
\label{eq:trace}
\end{equation}
\[
\ \frac{\left( m_{3}^{D}\,^{2}{\rm c}^{2}+
m_{2}^{D}\,^{2}{\rm s}^{2}\right)
^{2}}{M_{3}\,^{2}}\ +\ \frac{2(m_{3}^{D}\,^{2}-
m_{2}^{D}\,^{2})^{2}{\rm c}^{2}
{\rm s}^{2}\,{\cos 2\delta }}{M_{2}\,M_{3}} \ , 
\] 
where $\theta $, $\delta$ are defined in Eq.\ref{eq:umatrix}, 
${\rm c}=\cos \theta ,
\ {\rm s}=\sin \theta $. Note that the phases $\alpha_{2},
~\alpha_{3}$ in Eq.\ref{eq:mass} cancel out in these constraints 
and, thus, remain undetermined.

The mass matrix $m$ is diagonalized by a unitary rotation 
$V$ on $\nu$ 's: 
\begin{equation}
V\ =\ \left( 
\begin{array}{cc}
e^{i\beta _{2}} &  \\ 
& e^{i\beta _{3}}
\end{array}
\right) \left( 
\begin{array}{cc}
{\cos }\varphi & {\sin }\varphi \,e^{-i\epsilon } \\ 
-{\sin }\varphi \,e^{i\epsilon } & {\cos }\varphi
\end{array}
\right) \ ,
\label{eq:unit} 
\end{equation}
where $0\leq \varphi \leq \pi /2\,\ ,\ 0\leq \epsilon <\pi $. 
The `Dirac' mixing matrix in Eq.\ref{eq:dirac} is now multiplied 
by $V^{\dagger }$ on the left and, after phase absorptions, 
takes the form 
\begin{equation}
\left( 
\begin{array}{cc}
{\cos }\theta _{23} & {\sin }\theta _{23}\,e^{-i\delta _{23}} \\ 
-{\sin }\theta _{23}\,e^{i\delta _{23}} & {\cos }\theta _{23}
\end{array}
\right) \ ,\ 
\label{eq:mixing}
\end{equation}
where $0\leq \theta _{23}\leq \pi /2\,\ ,\ 0\leq 
\delta _{23}<\pi $. Here, 
$\theta _{23}$ (or $\theta _{\mu \tau }$) is the physical mixing 
angle in the $2-3$ leptonic sector and can be determined from the 
requirement that its cosine is equal to the 
modulus of the complex number 
${\cos }\varphi \,{\rm {\cos }\theta }^{D}+
{\rm {\sin }\varphi \sin \theta }
^{D}\,e^{i(\xi -\epsilon )}\ ,{\rm \ }$
where $-\pi \leq \xi -\epsilon =\beta _{2}-\beta _{3}-\epsilon 
\leq \pi $. The phases $\beta_{2}$, $\beta_{3}$ and 
$\xi$ remain undetermined due to the arbitrariness of 
$\alpha_{2}$, $\alpha_{3}$. Thus, the precise value of 
$\theta _{23}$ cannot be found. However, we can determine the range 
in which $\theta _{23}$ lies: 
$|\,\varphi -\theta ^{D}|\leq \theta _{23}
\leq \varphi +\theta ^{D},\ {\rm {
for}\ \varphi +\theta }^{D}\leq \ \pi /2\,\cdot$

\par
We now need the asymptotic values of $m^{D}_{2,3}$ ,~$\theta^{D}$. 
Approximate $SU(4)_{c}$- invariance in the up quark 
and neutrino sectors gives $m^{D}_{2}=m_{c}$,~$ m^{D}_{3} = m_{t}$, 
sin$\theta^{D}= \mid V_{cb}\mid$ `asymptotically'.
Renormalization of light neutrino masses and mixing between $M_{X} $ 
and $ M_{Z}$ is also included~\cite{blp,bs} assuming the MSSM spectrum 
and large tan$\beta \approx m_{t}/m_{b}$. In the framework of 
`hierarchical' light neutrino masses $(m_{3} \gg m_{2} \gg m_{1})$, 
the small angle MSW resolution of the solar neutrino puzzle implies~\cite{s} 
$1.7 \times 10^{-3}$ eV $\stackrel{_<}{_\sim}
m_{2} \stackrel{_<}{_\sim} 3.5 \times 10^{-3}$ eV. Finally, 
$m_{3}$ is restricted by the cosmological bound 
$m_3 \stackrel{_<}{_\sim}$ 23 eV (for $h \approx 0.5$).

\par
We are now ready to derive~\cite{lss,neu} useful restrictions on 
$M_{2,3}$. Assume that both $M_{2,3} \leq m_{infl}/2$. 
Then the inflaton predominantly decays to the heaviest of the two. 
The determinant condition implies that the lowest possible value of 
the heaviest $M_{2,3}$ is about $10^{11}$ GeV giving 
$T_r \stackrel{_>}{_\sim} 10^{10}$ GeV, in conflict with the 
gravitino constraint~\cite{gravitino} (see Sec.\ref{sec:reheathybrid}). 
So we are obliged to require that $1.72 \times 10^{13}$ GeV 
$\approx m_{infl}/2 \leq M_{3} \stackrel{_<}{_\sim} 2.5 
\times 10^{13}$ GeV, where the upper bound comes from the requirement 
that the coupling constant of the non-renormalizable term responsible for 
the mass of the heaviest $\nu^{c}$ does not exceed unity. 
This requirement also implies that $x_{Q} \stackrel{_<}{_\sim} 2.6$, 
which makes our choice in Sec.\ref{subsec:susy}, $x_{Q} \approx 2$, 
central. In summary, we see that 
i) $M_{3}$ is constraint in a narrow range, and ii) the inflaton 
decays to the second heaviest right handed neutrino, $\nu^{c}$, 
with mass $M_{2}$.

\par
Baryons can be produced, in the present scheme, only via a primordial 
leptogenesis from the decay of $\nu^{c}$ 's emerging as decay
products of the inflaton (see Sec.\ref{subsec:leptons}). The  lepton 
asymmetry is then partially converted into the observed baryon 
asymmetry of the universe by `sphaleron' effects as we explained in 
Sec.\ref{subsec:sphaleron}. From Eq.\ref{eq:genlept}, the lepton 
asymmetry is
\begin{equation}
\frac{n_{L}}{s}=\frac{9\,T_{r}}{8\pi \,m_{infl}}\,\frac{M_{2}}
{M_{3}}\,\frac{{\rm c}^{2}{\rm s}^{2}\ \sin 2\delta \ 
(m_{3}^{D}\,^{2}-m_{2}^{D}\,^{2})^{2}}{|\langle 
h^{(1)}\rangle|^{2}(m_{3}^{D}\,^{2}\ {\rm s}^{2}\ 
+\ m_{2}^{D}\,^{2}{\rm \ c^{2}})}\ \cdot
\label{eq:lepton}
\end{equation}
Renormalization effects should also be included in this formula. 
Assuming the MSSM spectrum between 1 TeV and $M_{X}$, we saw in 
Sec.\ref{subsec:sphaleron} that the observed baryon asymmetry 
$n_{B}/s$ is related to $n_{L}/s$ by $n_{B}/s =(-28/79) (n_{L}/s)$.

\begin{figure}
\psfig{figure=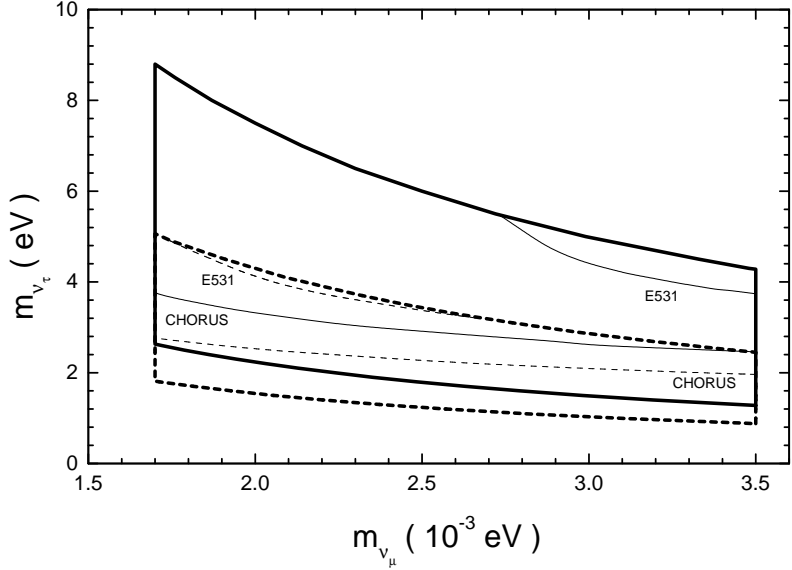,height=3in,angle=-90}
\caption{The allowed regions in the $m_{\nu_{\mu}}$,
~$m_{\nu_{\tau}}$ plane.
\label{fig:mass}}
\end{figure}

\par
We will now extract~\cite{neu} restrictions on light neutrino masses 
and mixing. Take a specific value of $M_{3}$ (in practice, 
we take its two extreme values $m_{infl}/2 $ or $2.5 \times 
10^{13}$ GeV). For any pair $m_{2}$,~$m_{3}$, we use the determinant 
condition to evaluate $M_{2}$ and, thus, $T_{r}$. 
The gravitino constraint~\cite{gravitino} 
($T_{r} \stackrel{_<}{_\sim} 10^{9}$ GeV)
then gives a lower bound in the $m_{2}$, $m_{3}$ plane. This
bound together with the small angle MSW restriction~\cite{s} on $m_{2}$ yields 
a lower bound on $m_{3}$, namely $m_{3} \stackrel{_>}{_\sim} 0.9 $ 
eV (for $M_{3}=2.5 \times 10^{13}$ GeV) or 1.3 eV 
(for $M_{3}=m_{infl}/2$). 
The trace condition is solved with respect to  $\delta = 
\delta (\theta)$, $0 \leq \theta \leq \pi/2$,  which is then 
substituted in Eq.\ref{eq:lepton} to yield $n_{L}/s = (n_{L}/s) 
(\theta)$. Imposing the `low' deuterium bound~\cite{deuterium} 
on $n_{B}/s$ (0.02 $\stackrel{_<}{_\sim} \Omega_{B}h^{2} 
\stackrel{_<}{_\sim}$ 0.03), we find the range 
of $\theta$ where this bound is satisfied. If such a range exists, 
we keep $m_{2}$,~$m_{3}$ as satisfying the baryogenesis
constraint. This gives an upper bound in the $m_{2}$,~$ m_{3}$ plane 
which together with the MSW restriction on $m_{2}$ yields an upper bound 
on $m_{3}$, namely $m_{3} \stackrel{_<}{_\sim} 5.1$ eV (for $M_{3}
=2.5 \times 10^{13}$ GeV) or 8.8 eV (for $M_{3}=m_{infl}/2$). 
The allowed area in the $m_{2}$,~$m_{3}$ plane is 
depicted~\cite {neu} in Fig.\ref{fig:mass}, where the thick solid
(dashed) line corresponds to $M_{3}=m_{infl}/2$ ($M_{3} =2.5 
\times 10^{13}$ GeV). The overall allowed range for $m_{\nu_{\tau}}$ 
is~\cite{neu} 1 eV $\stackrel{_<}{_\sim} m_{\nu_{\tau}} 
\stackrel{_<}{_\sim}$ 9 eV, which is interesting since the value of 
$m_{\nu_{\tau}}$ required by the cold plus hot dark matter 
scenario~\cite{chdm} for large scale structure formation in the 
universe is centrally located in this range.

\begin{figure}
\psfig{figure=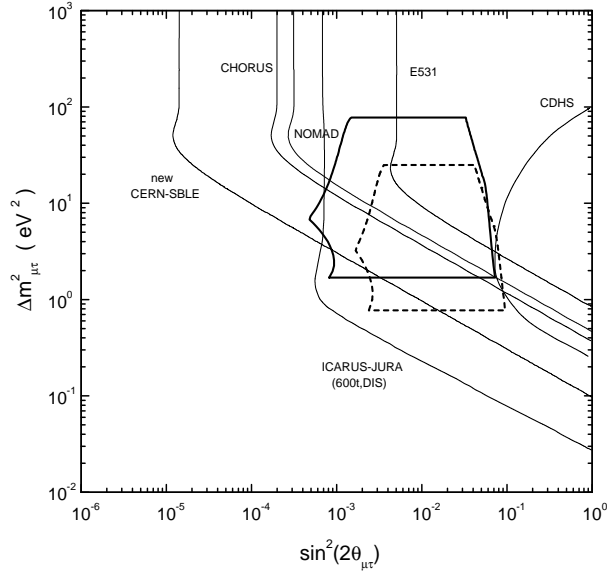,height=3in}
\caption{The allowed regions in the $\nu_{\mu}$-$\nu_{\tau}$ 
oscillation plot.
\label{fig:oscillation}}
\end{figure}

\par
The mixing angle $\theta_{\mu \tau}$ can also be restricted. 
For every allowed $m_{2}$,~$m_{3}$ pair and every 
$\theta$ satisfying the baryogenesis constraint, we
construct $V$ in Eq.\ref{eq:unit} and, consequently, $\varphi,
~\epsilon$ and the allowed range of $\theta_{\mu \tau}$,~$\mid 
\varphi - \theta^{D} \mid \leq \theta_{\mu \tau} \leq \varphi + 
\theta^{D}$. The union of all these ranges for all allowed $\theta$ 's 
and $m_{2}$ 's for a given $m_{3}$ gives the range of 
$\theta_{\mu \tau}$ which is allowed for this value of $m_{3}$. 
All these ranges for all allowed $m_{3}$ 's constitute the allowed area
on the oscillation diagram, which is depicted~\cite{neu} in 
Fig.\ref{fig:oscillation} (notation as in Fig.\ref{fig:mass}) in 
confrontation to past, ongoing and planned experiments. The central part 
of this allowed area will be tested by the ongoing Short Baseline 
Experiments (SBLE) at CERN, NOMAD/CHORUS.
Possibly negative result from NOMAD/CHORUS will exclude a significant 
part of the allowed domains in Figs.\ref{fig:mass},
\ref{fig:oscillation} reducing the upper bound on the tau-neutrino 
mass, $m_{\nu_{\tau}}$, to 3.7 eV. The new CERN SBLE(TOSCA) together 
with the new CERN Medium Baseline Experiment (MBLE) ICARUS-JURA (600t,DIS) 
will cover all our predicted area on the oscillation diagram.

\par
In summary, hybrid inflation, baryogenesis and neutrino oscillations 
have been linked in the context of a supersymmetric model based on a 
left-right symmetric gauge group. Our scheme leads to stringent 
restriction on $m_{\nu_{\tau}}$ and $\theta_{\mu \tau}$ to
be tested by ongoing and planned experiments. These restrictions 
are derived by mainly `physical' arguments
(gravitino and baryogenesis constraints) supplemented by a `minimal' 
input from fermion mass matrix ansaetze (only 3 input parameters ) 
and experiments (MSW resolution of the solar neutrino problem).
The choice of the gauge group is crucial since, in this case,
$\phi$ has  the quantum numbers of $\nu^{c}$ and , thus, decays to 
$\nu^{c}$~'s producing an initial lepton asymmetry. As a consequence, 
the gravitino and baryogenesis constraints restrict the neutrino 
parameters.

\section*{Acknowledgment}
This work was supported in part by the European Commission under the 
Human Capital and Mobility programme, contract number CHRX-CT94-0423.

\end{document}